\documentclass[a4paper,11pt]{article}
\pdfoutput=1

\usepackage{jheppub} 

\usepackage[T1]{fontenc} 
\usepackage[dvipsnames]{xcolor}

\usepackage{booktabs}
\usepackage{float}
\usepackage{titlesec}
\usepackage{capt-of}
\usepackage{slashed,bm}

\usepackage{mathrsfs}
\usepackage{amsmath}
\usepackage{tikz}
\usepackage[com pat=1.1.0]{tikz-feynman}
\usetikzlibrary{calc}
\usepackage{slashed}
\usepackage{MnSymbol}
\usepackage{dsfont}
\usepackage{mathtools}
\usepackage{wrapfig}
\usepackage{bbold}

\newcommand{\dd}{\textmd{d}}
\newcommand{\Z}{\mathcal{Z}}

\DeclarePairedDelimiter\bra{\langle}{\rvert}
\DeclarePairedDelimiter\ket{\lvert}{\rangle}
\DeclarePairedDelimiterX\braket[2]{\langle}{\rangle}{#1\,\delimsize\vert\,\mathopen{}#2}

\newcommand{\Omegacl}{\Omega_{\rm macr}}
\newcommand{\chimumu}{c_2}
\newcommand{\xiW}{\xi_{\rm W}}
\newcommand{\xiS}{\xi_{\rm S}}
\newcommand{\mubar}{\bar\mu} 
\newcommand{\Leff}{\mathscr{L}_{\rm eff}}
\newcommand{\iE}{\hat E}
\newcommand{\imu}{\hat \mu}
\newcommand{\nf}{n_F}
\newcommand{\ff}{f_F}
\newcommand{\Nf}{N_F}
\newcommand{\nb}{n_B}
\newcommand{\Sb}{S_B}

\title{On electric fields in hot QCD: infrared regularization dependence}

\renewcommand*{\thefootnote}{\fnsymbol{footnote}}
\author[a]{G. Endr\H{o}di,}
\author[a]{G. Mark\'o,}
\author[b]{and L. Sandbote\footnote{Corresponding author}}

\affiliation[a]{Institute of Physics and Astronomy,
ELTE E\"otv\"os Lor\'and University,\\
P\'azm\'any P.\ s\'et\'any 1/A, H-1117 Budapest, Hungary}
\affiliation[b]{Department of Physics and Helsinki Institute of Physics,\\
P.O.~Box 64, FI-00014 University of Helsinki, Finland}

\emailAdd{gergely.endrodi@ttk.elte.hu}
\emailAdd{marko.gergely@ttk.elte.hu}
\emailAdd{leon.sandbote@helsinki.fi}

\abstract{
We study the impact of background electric fields on a hot plasma of charged particles -- a setting relevant for the early stages of heavy-ion collisions as well as laser pulse experiments.
Historically, the electric susceptibility -- encoding the behavior of the hot medium for weak fields -- has been defined within two different formalisms, leading to two distinct results at nonzero temperature. With the help of an exact fermion propagator in a homogeneous electric background field at nonzero temperature and finite volume on the one hand, and an improved perturbative result on the other, we identify the origin of this disagreement. The equilibrium conditions for the system are discussed and the role of the thermodynamic ensemble used to describe the system is highlighted. Finally, we construct the electric susceptibility in a simplified hadron resonance gas model, relevant for the strongly interacting medium in the low-temperature regime.}

\keywords{thermal field theory, background electromagnetic fields, linear response}

\begin{document}
\maketitle
\flushbottom

\renewcommand*{\thefootnote}{\arabic{footnote}}

\section{Introduction}

A key feature of a quantum field theoretical system is its response 
to classical background fields. With regard to the electromagnetic response, the background field couples to the electrically charged particles originating from virtual or thermal fluctuations, forming a medium with non-trivial magnetization and polarization. Two prime examples for such a behavior are the thermal media described by Quantum Chromodynamics (QCD) and Quantum Electrodynamics (QED). 

Here we focus on the electric polarization of the thermal medium, induced by background electric fields. For QCD matter, this response is relevant for the behavior of the strongly interacting medium formed in the early stages of heavy-ion collision experiments~\cite{Deng:2014uja,Voronyuk:2014rna,Huang:2015oca}. In turn, the reaction of the QED plasma to strong electric fields is put to the test routinely in laser pulse experiments~\cite{Gies:2008wv,DiPiazza:2011tq,Fedotov:2022ely}. 
Both setups give rise to a broad range of fascinating physical phenomena~\cite{Hattori:2023egw}.

While QED interactions are essentially perturbative, the QCD medium is strongly coupled at low to intermediate energies and may only be studied via non-perturbative first principles methods like lattice QCD simulations. Nevertheless, leading-order perturbation theory was observed to describe the response of the QCD medium to weak background fields surprisingly well. In particular, a quantitative agreement between one-loop perturbative predictions and lattice QCD simulations was observed down to temperatures of the order of $T\approx 300\textmd{ MeV}$, both for weak magnetic~\cite{Bali:2014kia,Bali:2020bcn} and weak electric fields~\cite{Endrodi:2023wwf}, see also the recent review~\cite{Endrodi:2024cqn}.
Perturbative QCD calculations are in general expected to become valid at high energies, where the QCD coupling decreases due to asymptotic freedom. Altogether, perturbation theory turns out to be a reasonable tool for both QCD and QED plasmas in this context and in this paper we consider a perturbative treatment of these setups to one loop order.

In addition, the generalization of the QED result to constituent particles (i.e.\ hadrons) also allows to predict the electromagnetic response of QCD matter at low temperatures, where it is well approximated by hadron gas models~\cite{Hagedorn:1965st,Karsch:2003vd}. In fact, such a hadron resonance gas model has been constructed for the case of magnetic fields~\cite{Endrodi:2013cs,Bali:2020bcn} and has been revisited recently~\cite{Vovchenko:2024wbg,Marczenko:2024kko,Huovinen:2025nwl}. Here we work out an analogous model for electric background fields.

The response of the medium is encoded by the behavior of the thermodynamic potential $\Omega$. The leading-order coefficient, in a weak field expansion, is the electric susceptibility
\begin{equation}
 \xi=-\left.\frac{\mathrm{d}^2 \Omega}{\mathrm{d} (eE)^2}\right|_{E=0}\,,
 \label{eq:defxi}
\end{equation}
 which will be in the focus of our discussion. It is normalized by the elementary electric charge $e>0$ for convenience.\footnote{In this manner, multiplicative renormalization constants do not appear in $\xi$, see e.g.\ Ref.~\cite{Endrodi:2024cqn}.}
As we will discuss in detail below, the response of the thermal system to the electric field is twofold. First, it involves a {\it macroscopic} rearrangement of electric charge carriers, generating an inhomogeneous charge distribution. Second, it includes the {\it quantum} response of the medium, encoded in the propagation of particles.
In the present paper, we are interested in the latter contribution and will therefore carefully isolate the two responses in order to compute the susceptibility.

Interestingly, the susceptibility~\eqref{eq:defxi} can be calculated via two fundamentally different approaches. One possibility is to calculate the thermodynamic potential $\Omega(E)$ using Schwinger's exact fermion propagator~\cite{Schwinger:1951nm} at nonzero electric field, followed by a weak field expansion to obtain~\eqref{eq:defxi}. This strategy was followed both using real-time propagators~\cite{Loewe:1991mn,Elmfors:1994fw,Elmfors:1998ee} as well as within the 
imaginary time formalism~\cite{Gies:1998vt,Gies:1999xn,PhysRevD.108.076002} 
at arbitrary temperatures and fermion masses. The so defined susceptibility will be denoted by $\xiS$.
Alternatively, one may expand $\Omega$ in a general background photon field to express the susceptibility directly
in terms of the photon vacuum polarization diagram at $E=0$. This 
approach was employed in the seminal paper by Weldon~\cite{Weldon:1982aq} for massless fermions and generalized to the massive case in~\cite{Endrodi:2022wym}, see also~\cite{Carignano:2017ovz,Ferreira:2023cqw}. This determination of~\eqref{eq:defxi} will be denoted by $\xiW$. As first pointed out in~\cite{Endrodi:2022wym}, the two approaches surprisingly disagree,
\begin{equation}
\begin{split}
    \xiS &= \,-\frac{1}{6\pi^2}\int_m^\infty
    \frac{\dd\omega}{\sqrt{\omega^2-m^2}} \, \Bigl[ 2n_F(\omega) + \omega\,n_F'(\omega) \Bigl] \\
    \xiW &= \,-\frac{1}{6\pi^2}\int_m^\infty
    \frac{\dd\omega}{\sqrt{\omega^2-m^2}} \, \Bigl[ 2n_F(\omega) - \omega\,n_F'(\omega) \Bigl],
\end{split}
\end{equation} 
where $n_F(\omega) = 1/(e^{\beta\omega}+1)$ is the Fermi-Dirac distribution and $m$ the mass of the fermion. 
In particular, the difference of the two susceptibilities in a high-temperature expansion reads $\xiS-\xiW=1/(6\pi^2)+\mathcal{O}(m^2/T^2)$. Finally, we also mention the recently developed representation of the real-time formalism to treat background electric fields~\cite{Fukushima:2025eyt}.

This disagreement poses a substantial discrepancy between the above two seminal methods. Schwinger's approach forms the basis for Euler-Heisenberg effective actions employed for QED thermodynamics and in QCD effective models~\cite{Miransky:2015ava} but is also highly important for laser physics~\cite{Hattori:2023egw}. In turn, Weldon's approach, generalized to color interactions, has been influential for hard thermal loop 
perturbation theory~\cite{Braaten:1991gm,Blaizot:2001nr} and for quark-gluon plasma physics, see e.g.\ Ref.~\cite{Arnold:2003zc}.
Besides the electric susceptibility, the weak-field response at nonzero temperature is also relevant for other observables and the phase diagram in the temperature-electric field plane~\cite{Yamamoto:2012bd,Tavares:2019mvq,Loewe:2021mdo,Endrodi:2021qxz,Yang:2022zob,Endrodi:2023wwf,Tavares:2023ybt,Tavares:2024edx}.
It is therefore essential to identify the reason for $\xiS\neq\xiW$ and to resolve this discrepancy.

Previously, the disagreement was argued to be rooted in the infrared (IR) divergence that appears due to the singular equilibrium charge profile in homogeneous electric fields in an infinite volume~\cite{Endrodi:2022wym}.\footnote{We mention that infrared divergences also show up in calculations of the electric permittivity in the vacuum~\cite{Taya:2023ltd}.} Moreover, the ambiguity was conjectured to be related to the ordering of infrared limits like the thermodynamic limit and the weak electric field limit~\cite{Endrodi:2022wym}. In this paper, we address this conjecture and explicitly demonstrate that the reason for $\xiS\neq\xiW$ is indeed rooted in infrared physics,
but emerges rather because the global spatial averaging involved in $\xi$ and the limit of homogeneous electric fields do not commute. Thus, the settings in e.g.\ Ref.~\cite{Weldon:1982aq} and Ref.~\cite{Elmfors:1994fw} indeed correspond to different physics, i.e.\ to observables with different measurement instructions.
Furthermore, we point out the importance of using different thermodynamic ensembles -- the local grand canonical ensemble, where density fluctuations around the equilibrium density profile are allowed, and the local canonical ensemble, where the density profile is fixed. 

This paper is structured as follows: 
In Sec.~\ref{sec:eqthermo}, we discuss the equilibrium at finite temperature modified by the presence of an electric background field and revisit the IR divergence that is caused by $E\neq0$. This is followed by the introduction of an IR regularization within both a canonical and a grand canonical thermodynamic ensemble. The possible ordering of the various relevant infrared limits and the resulting values of the susceptibility are summarized in Sec.~\ref{sec:mainresult}. 
The calculations to obtain the susceptibilities are split up into Secs.~\ref{sec:3} and~\ref{sec:4}. First, we analyze the interchangeability of the thermodynamic limit with the weak electric field limit and second, that of the global spatial averaging with the limit of homogeneous fields.  
We conclude the discussion by presenting an electric susceptibility of a hadron resonance gas model and comparing it to lattice QCD results of Ref.~\cite{Endrodi:2023wwf}. Based on the infrared limits involved, we examine the physical interpretation of the two electric susceptibilities which are at tension with each other in the conclusion.   
In the Appendices we present further technical details:
the fermion propagator in the presence of a background electric field in a finite volume and nonzero temperature in App.~\ref{app:1}, the local charge and current density based on this propagator in App.~\ref{app:2}, various cross-checks in App.~\ref{app:3} as well as detailed calculations in Apps.~\ref{app:4} and \ref{app:5}.
Some calculations and results presented in this paper have originally been developed by the authors in Ref.~\cite{Leon_thesis} and are extended and further clarified here.

\section{Equilibrium and thermodynamic ensembles}
\label{sec:eqthermo}

For concreteness, throughout this paper we use Euclidean space-time coordinates. The metric is $\delta_{\mu\nu}$, lower indices of coordinates $x_i$, $x_4$ are Euclidean coordinates and for the integration measure $\dd^4x=\dd x_4\dd^3x$ holds. However, the relation $ix_0=x_4$ for the real time $x_0$ will be applied in the context of analytic continuations between real and imaginary electric fields. For the gauge potential, this relation implies $A_4 =-i A_0 $. 

Having set the stage, we consider a quantum field theory with fermionic degrees of freedom $\psi$, having mass $m$ and electric charge $e$ at nonzero temperature $T$.
The thermodynamic potential $\Omega$ for such fermions (in an external gauge field $A_\nu$ but without dynamical photons) takes the form,
\begin{equation}\label{eq:defLeff}
 \Omega = -\frac{T}{V}\log \Z = -\frac{T}{V}\log\det\left[\slashed{D}(A)+m\right] = -\frac{T}{V} \int \dd^4 x \int \dd m\, \textmd{Tr}\, G(x,x)
 =\frac{T}{V}\int \dd^4 x\, \Leff(x)\,,
\end{equation}
where we differentiated and integrated with respect to the fermion mass, resulting in the appearance of the traced fermion propagator $G$ (over a closed loop starting and ending at the coordinate $x$) for the specific background field $A_\nu$.
The last equation defines $\Leff(x)$, the local effective Lagrangian in a Euclidean sense. 

Consider a classical picture. The presence of a background electric field $E$ in the direction $x_3$ exerts the Lorentz force on charged particles, accelerating them in the $x_3$ direction.
The local charge density distribution $n(x_3)$ that forms after equilibration will therefore be inhomogeneous. 
Notice that -- since the electric field is only external and there are no dynamical photons in our system -- the displaced charges themselves do not produce an internal electric field.
The charge distribution is fixed by the requirement that pressure gradients (i.e.\ diffusion forces due to fermionic repulsion) and the Lorentz force cancel each other exactly and can therefore lead to an equilibrium~\cite{Endrodi:2022wym}. 
For an arbitrary potential $A_0(x_3)$, it can be shown that this density profile $n(x_3)$ is generated by an effective chemical potential~\cite{landau1995electrodynamics,Luttinger:1964zz}
\begin{equation}
    \mubar(x_3) = \mu + eA_0(x_3)
    \,,
    \label{eq:mubar}
\end{equation}
where $\mu$ controls the total amount of charge in the system. The behavior $\mubar(x_3\to\pm\infty)\to\mp\infty$ or $n(x_3\to\pm\infty)\to\mp\infty$, 
for an unbounded potential $A_0(x_3)$ towards infinity (e.g.\ for a homogeneous electric field $A_0=-Ex_3$) has the consequence of a singular global charge distribution, creating an IR divergence present in all global thermodynamic observables and rendering the thermodynamic potential in Eq.~(\ref{eq:defLeff}) ill-defined.
At the root of this IR divergence is the impossibility of creating or measuring an infinite volume system, completely permeated by a homogeneous electric field and forces us to consider a thermodynamic system which only extends over a finite distance $L_3$.

The macroscopic charge distribution gives a contribution to $\Omega$ -- and, to $\xi$ -- that
originates purely from the local charge density. To assess this contribution, consider a subsystem around the coordinate value $x_3$ in the direction of the electric field. The charge density in this subsystem follows from $\mubar(x_3)$ in Eq.~(\ref{eq:mubar}). Neglecting 
quantum effects of the electric field on the particles, the associated thermodynamic potential reads
\begin{equation}\label{eq:omegafree}
    \left.\Omegacl(\mu)\right|_{\mu=\mubar(x)}=\Omegacl(0) -\frac{\chimumu}{2}\cdot(\mu+eA_0(x))^2 +\mathcal{O}(\mubar(x)^4)\,,
\end{equation}
where $\Omegacl(\mu)$ stands for the $E=0$ thermodynamic potential as a function of the chemical potential and indicates the macroscopic effect of the charge profile.
Here we carried out a Taylor-expansion in the chemical potential, introducing the fermion charge susceptibility $\chimumu=- \partial^2_\mu\,\Omegacl$ as the leading-order coefficient. As~\eqref{eq:omegafree} reveals, there are terms of $\mathcal{O}(A_0^2)$ i.e.\ terms of $\mathcal{O}(E^2)$ that contribute to~\eqref{eq:defxi}, which originate purely from this macroscopic rearrangement of electric charges. For a globally neutral medium, $\mu=0$ and this contribution can be quantified solely by $\chimumu$.
We are not interested in this macroscopic contribution and would like to isolate it from the quantum response of the medium to the electric field 
and ultimately construct the associated susceptibility $\xi$.\footnote{We note that whether or not the contribution of the inhomogeneous charge distribution is to be taken into account in the electric response of a certain system depends on the actual phenomenological application. The time-scales of the different types of responses: the redistribution of electric charges on the one hand and the local quantum behavior of the medium on the other are inherently different. We furthermore note that the high-temperature expansion of the two contributions also differ: $\chimumu\sim T^2$ dominates over the quantum response $\xiS,\,\xiW\sim -\log(T)$, stressing the potential importance of the former term at high $T$.} To that end, we proceed as follows.

First, let us consider a system in a spatial volume with length $L_3$ in the direction of the electric field. 
This system can be split up into subvolumes of length $\ell$, which are already decorrelated from each other,
so that~\eqref{eq:defLeff} is the sum of the thermodynamic potentials for the subvolumes.
For one subvolume
centered around the point $y_3$, the 
thermodynamic potential reads,
\begin{equation}
 \Omega_{\ell}(y_3)=\frac{1}{\ell}\int_{y_3-\ell/2}^{y_3+\ell/2}\dd x_3\,\Leff(x_3) =  -\frac{1}{\ell}\int_{y_3-\ell/2}^{y_3+\ell/2}\dd x_3 \int \dd m \, \textmd{Tr}\, G(x,x).
 \label{eq:leff1}
\end{equation}

Second, considering $2N+1$ subsystems of length $\ell = L_3/(2N-1)$, we modify the $j$-th subsystem by a chemical potential shift $-eA_0(j\,\ell)$ and arrange them in a heat bath next to each other. The thermodynamic potential thus reads
\begin{equation}
    \Omega_N(E,L_3) = \sum_{j=-N}^N\, \Bigl[ \Omega_\ell(y_3 = j\,\ell)\Bigl]_{\mu\,\rightarrow\, \mu-eA_0(j\,\ell)}.
\end{equation}
Alternatively, we arrange similar smaller systems next to each other but place them in a thermodynamic environment where particle exchange is forbidden. This is achieved via a local Legendre transform that replaces the chemical potential $\mu(n)$ by the charge density $n$, resulting in the free energy density 
\begin{equation}
    f_N(E,L_3) = \sum_{j=-N}^N\, \Bigg[ \Omega_\ell(y_3 = j\,\ell) + \mu \frac{\partial\Omega(y_3 = j\,\ell)}{\partial\mu} \Bigg]_{\mu\,\rightarrow\, \mu(n)-eA_0(j\,\ell)}.
\end{equation}

Third, we perform the $N\to\infty$ limit while keeping $L_3$ fixed. This limit eliminates the contribution purely due to the inhomogeneous charge distribution, since the chemical potential shift $\mu(n)-eA_0(j\,\ell)$ sets the effective chemical potential of Eq.~(\ref{eq:mubar}) to a constant.

In the following, we consider a globally neutral system.
We arrive at our grand canonical approach allowing for particle exchange,
\begin{equation}\label{eq:Omegaloc}
    \Omega(E,L_3)=\,\frac{1}{L_3}\int \mathrm{d}x_3\  \Leff(x_3,E,\mu)\Bigl|_{\mu=-eA_0(x_3)}.
\end{equation}
While our canonical approach with a fixed amount of charge yields
\begin{equation}
   f(E,L_3)= \,\frac{1}{L_3} \int \mathrm{d}x_3\, \left[ \Leff(x_3,E,\mu) +\mu \frac{\partial\Leff(x_3,E,\mu)}{\partial\mu} \right]_{\mu=-eA_0(x_3)},
   \label{eq:floc}
\end{equation}
which is related to the former approach via a Legendre transform on the level of the local effective Lagrangian.

Now, a small change in the background electric field modifies the above thermodynamic potential $\Omega$ only by affecting charged particles locally without the response of the inhomogeneous charge distribution. Notice that in this formalism, the effective chemical potential $\bar\mu$ of Eq.~\eqref{eq:mubar}, appearing in $\Leff$ becomes a constant, independent of $x_3$ and $E$. Still, $\Leff$ depends explicitly on the electric field through the fermion propagator. 
This latter dependence quantifies the quantum response of the medium to the electric field, isolated from the impact of the rearranged charge distribution,  Eq.~(\ref{eq:omegafree}).
The electric susceptibility~\eqref{eq:defxi} we consider measures the quantum effect. 
Clearly we can define the susceptibility also in the canonical ensemble, by replacing $\Omega$ by $f$ in~\eqref{eq:defxi}.

The thermodynamic potential Eq.~(\ref{eq:Omegaloc}) has already been considered in Refs.~\cite{Elmfors:1994fw,Elmfors:1998ee}, while the potential of Eq.~(\ref{eq:floc}) was used in Ref.~\cite{Endrodi:2022wym} to calculate the electric susceptibility. Both of these approaches relied on setting the effective chemical potential of Eq.~(\ref{eq:mubar}) to a constant value. 

Finally, we note that the thermodynamic potentials and the effective Lagrangian contain ultraviolet divergent terms besides pure vacuum fluctuations, which are eliminated via electric charge renormalization, or, equivalently, the renormalization of the vector potential $A_\mu$~\cite{Schwinger:1951nm}. Since this issue is well known (see~\cite{Endrodi:2024cqn} for a recent review) and is not central to the present discussion, we keep the corresponding details implicit.

\subsection{Infrared regularizations}\label{sec:mainresult}

As already discussed, homogeneous electric fields in the infinite volume imply an unbounded behavior for the equilibrium charge density towards $x_3\to\pm\infty$ and cause infrared divergences in spatial integrals. The thermodynamic potentials~\eqref{eq:Omegaloc} and~\eqref{eq:floc} are, however, constructed in such a way that the impact of the macroscopic charge distribution is eliminated in them, rendering both $\Omega(E)$ and $f(E)$ infrared finite. As we will see, this occurs in a delicate manner: IR divergences show up both in the explicit dependence on $E$ as well as in the implicit dependence via $\mu(E)$ and the two cancel each other exactly.

In order to avoid these divergences in all intermediate steps of the calculation, we work with a bounded $eA_0(x_3)$ function and to do so, we impose an IR regularization. 
The final result for the electric susceptibility describing the response to a homogeneous field in infinite volume is then achieved by removing the IR regularization.

We consider the two possible IR regularizations.
First, we constrain the size $L_3$ of the system in the direction of the electric field to be finite. To be specific, the coordinate origin is placed in the center, i.e.\ $-L_3/2\le x_3\le L_3/2$ and the volume average is performed by an integration over the volume while normalizing by the volume, i.e.\ for the finite volume case $\frac{1}{L_3}\int\dd x_3$ or $\lim_{L_3\to\infty}\frac{1}{L_3}\int\dd x_3$ for the infinite volume case. In this section, the volume average will be denoted simply by $\int\dd x_3$.
The susceptibility is calculated both in the canonical and grand canonical approach and the IR regularization is finally removed by taking the limit $L_3\to\infty$. 
Second, we deform the electric field
to an oscillatory field,
\begin{equation}\label{eq:finitekPotential}
    A_0(x_3) = - E\,\frac{\sin (kx_3)}{k}\,,
\end{equation}
with wavelength $2\pi/k$.
We again calculate $\xi$ both from $\Omega$ and from $f$. Anticipating one of our main results, in this case we will find that the spatial averaging involved in Eqs.~\eqref{eq:Omegaloc} and~\eqref{eq:floc} does not commute with the $k\to0$ limit. Such a non-commutation does not appear for the IR regularization involving the finite volume.

\begin{table}[h!]
\center
\begin{tabular}{c|c|c|c|}\label{tab:susceptable}
    & $\lim_{k\to0}\int \dd x_3$ & $\int \dd x_3 \lim_{k\to0}=\int \dd x_3 \lim_{L_3\to\infty}$ & $ \lim_{L_3\to\infty} \int \dd x_3$ \\ \hline
    canonical & $\xiW$ \cite{Endrodi:2022wym,Ferreira:2023cqw} & $\xiS$ &$\xiS$ \\ \hline
    grand canonical & $\xiS$ & $\xiS$~\cite{Loewe:1991mn,Elmfors:1994fw,Elmfors:1998ee,Gies:1998vt,Gies:1999xn} & $\xiS$ \\ \hline
\end{tabular}
\caption{The electric susceptibility $\xi$ in the local canonical and local grand canonical approaches, depending on the ordering of the spatial averaging $\int \dd x_3$ and the removal of the IR regularization (infinite wavelength limit $k\to0$ or infinite volume limit $L_3\to\infty$). We also indicate the references that employed the respective ensembles and orderings.}
\label{tab:susc}
\end{table}

Our results are summarized in Tab.~\ref{tab:susc}, which shows $\xi$ in the local canonical (top row) and the local grand canonical ensembles (bottom row). All possibilities, involving the removal of the IR regularizations $L_3$ or $k$ before or after spatial averaging, are indicated. (Note that before volume averaging, the infinite volume and infinite wavelength limits are identical by definition, indicated in the center column of Tab.~\ref{tab:susc}.)
In each case, the result is found to give either the value $\xiS$ as in Refs.~\cite{Loewe:1991mn,Elmfors:1994fw,Elmfors:1998ee,Gies:1998vt,Gies:1999xn} based on the Schwinger propagator, or $\xiW$ as in Refs.~\cite{Endrodi:2022wym,Ferreira:2023cqw} based on Weldon's gradient expansion.\footnote{\label{fn:factor2}
We further point out that when working with the oscillating electric field~\eqref{eq:finitekPotential}, an overall factor $1/2$ appears in the second electric field derivative in comparison to the homogeneous electric field case. This normalization factor affects the energy stored in the electric field in the same manner, being $\lim_{k\to0}\int \dd x_3\, [-\partial_{x_3}A_0(x)]^2/2 =\, E^2/4$ instead of $E^2/2$ and therefore cancels trivially from the susceptibility.
}
We also mention the studies~\cite{Weldon:1982aq,Carignano:2017ovz} following the approach with oscillatory fields, which would result in $\xiW$ when the response of the inhomogeneous charge profile is removed from the susceptibility.

All of our calculations rely on Eqs.~\eqref{eq:Omegaloc} and $\eqref{eq:floc}$ with the chemical potential set as $\mu=-eA_0(x_3)$ under the spatial integrals.
Specifically, we compute the cells of Tab.~\ref{tab:susc} involving $k\neq0$ as follows. The ordering $\lim_{k\to0}\int \dd x_3$ makes use of an expansion of the effective Lagrangian
for finite wavelength $2\pi/k$, followed by the spatial average $\int \dd x_3$ to arrive at the expansion of the thermodynamic potential.
The grand canonical ordering $\int \dd x_3 \lim_{k\to0}$ was already derived by the effective Lagrangian of Ref.~\cite{Elmfors:1994fw}, which is exact for homogeneous electric fields. As a further consistency check, we also rederive this result via a weak coupling expansion in Sec.~\ref{sec:4}.

The remaining cells of Tab.~\ref{tab:susc} are used to check the impact of using the finite volume or infinite volume effective Lagrangian. The ordering $\int \dd x_3 \lim_{L_3\to\infty}$ again relies on Ref.~\cite{Elmfors:1994fw} -- and agrees with $\int\dd x_3\lim_{k\to0}$ by construction. The ordering $ \lim_{L_3\to\infty} \int \dd x_3$ uses the exact finite volume effective Lagrangian for a homogeneous electric field, with an infinite volume limit in the end.
For the latter, we needed the fermion propagator in a finite volume, at nonzero temperature and electric field, which we derive in this paper for the first time.

As visible from Tab.~\ref{tab:susc}, a difference 
is only seen between the canonical and grand canonical ensembles in the case that the IR regularization is realized by oscillatory fields and the $k\to0$ limit is taken after spatial averaging (first column). Moreover, the $k\to0$ limit and the spatial averaging are found not to commute with each other in the canonical ensemble (first two cells in the top row). We get back to the interpretation of this finding and the physical difference between the oscillatory and homogeneous fields in the summary.
That the ordering of infrared limits is responsible for the mismatch $\xi_S\neq \xi_W$ has already been conjectured in Ref.~\cite{Endrodi:2022wym}. However, there the ordering of the weak field expansion $E\to0$ and the $L_3\to\infty$ limit was assumed to be at the root of the disagreement. We will investigate this issue further in Sec.~\ref{sec:3}.

\section{Electric field derivative and infinite volume limit}
\label{sec:3}

To address specifically the conjecture of Ref.~\cite{Endrodi:2022wym}, 
we compute the susceptibility using the two orderings of the limits $L_3\to\infty$ and $E\to0$ in the grand canonical ensemble. As we will demonstrate, we find the same value $\xi_S$ in both cases,
\begin{equation}\label{finitevolumesusceptibility}
     -\lim_{L_3\rightarrow\infty}\left.\frac{\mathrm{d}^2\Omega(E,L_3)}{\mathrm{d}(eE)^2}\right|_{E=0}
    =
     -\left.\frac{\mathrm{d}^2\Omega(E,L_3=\infty)}{\mathrm{d}(eE)^2}\right|_{E=0} = \xi_S \,,
\end{equation}
showing that this ordering of limits has no impact on the susceptibility.

To derive this result, we make use of the exact fermion propagator at finite temperature $T=1/\beta$ and finite length $L_3$. In order to treat an electric field in a finite volume with periodic boundary conditions, we need to consider an imaginary (Euclidean) electric background field $\iE=-iE$.
Accordingly, we take into account an imaginary chemical potential $\imu=-i\mu$. 
The propagator is derived by following the method of image sums, which was  also applied in Ref.~\cite{Adhikari:2023fdl} to scalar fields in the presence of homogeneous magnetic fields. 

The Euclidean vector potential satisfies $A_4(x)=-\iE x_3$ with all other components set to zero.
To realize this setup in a finite periodic volume, we make use of twisted spatial and antiperiodic temporal boundary conditions for the Dirac field (see e.g.\ the review~\cite{Endrodi:2024cqn}),
\begin{equation}
    \psi(x+L_3\hat{e_3})=e^{-i\mu_3L_3+ie\iE L_3x_4}\psi(x)\quad\text{and}\quad \psi(x+\beta\hat{e}_4)=-e^{i \imu\beta}\psi(x),
    \label{eq:bc}
\end{equation}
where $\hat{e}_\nu$ stands for the unit vector in the direction $\nu$ and the spatial `chemical potential' $\mu_3$ is introduced in order to control the charge current, to which we get back to below. 
Due to the twisted boundary conditions, the electric field is quantized as~\cite{Endrodi:2024cqn}
\begin{equation}
\label{eq:quantization}
    e\iE = \frac{2\pi}{\beta L_3}N_e,\qquad N_e\in\mathds{Z}. 
\end{equation}

The boundary conditions~\eqref{eq:bc} lead to constraints for the propagator $G$. 
The details regarding the derivation of the propagator and its precise form are included in App.~\ref{app:1}.
Based on our result for the propagator, the effective Lagrangian is found to be
\begin{equation}\label{efflag}
    \Leff(x_3) = 2 \int_0^\infty \frac{\mathrm{d}s}{s}\  e^{-sm^2}  \cosh(e\iE s)  \,g^\parallel_{L_3,\beta}(x_\parallel,x_\parallel|s) \, g^\bot(x_\bot,x_\bot|s)\,,
\end{equation}
where the coordinate is split into parallel $x_\parallel=(x_3,x_4)$ and perpendicular $x_\bot=(x_1,x_2)$ components. 
The functions $g^\parallel$ and $g^\bot$ can be found in Eqs.~\eqref{gfree},~\eqref{geven} and~\eqref{godd}.
Using the propagator, we also define the electric charge density and the corresponding current,
\begin{equation}
n(x) = \langle \bar\psi(x) \gamma^0 \psi(x)\rangle, \qquad j^i(x) = \langle \bar\psi(x) \gamma^i \psi(x)\rangle\,.
\label{eq:densitycurrent}
\end{equation}
Eq.~\eqref{efflag} depends on the coordinates via $g^\parallel_{L_3,\beta}$ implicitly through $\mubar(x)=\imu-e\iE x_3$ as well as through $\mubar_3(x)=\mu_3-e\iE x_4$.
This propagates into the coordinate-dependences of the charge density and of the current, which are visualized in App.~\ref{app:2}.

In Sec.~\ref{sec:eqthermo}, we argued that in order to isolate the response of the medium to the background electric field, we need to set $\bar\mu$ to a constant (zero for a globally neutral system) and, correspondingly $\imu=e\iE x_3$. Here we demonstrate the validity and impact of this choice in more detail.
If we keep $\imu$ a constant, we find the charge density $n(x)$ to depend on $x_3$, and after the space-time integral, $\Omega$ is in fact found to be independent of $\iE$ and of $T$. One might think about this choice as the situation, where the charged constituents of the medium have already been expelled from the bulk, rendering it equivalent to the vacuum. Instead, promoting $\imu$ to be $x_3$-dependent in the effective sense discussed in Sec.~\ref{sec:eqthermo}, such that $\mubar$ is a constant, we find that the $x_3$-dependence is eliminated and $\Omega$ depends on $\iE$ and $T$ as it should.
We encounter a similar situation with $\mu_3$. If we keep $\mu_3$ a constant, we find the current density $j^3(x)$ as well as the charge density $n(x)$ to depend on imaginary time $x_4$. Thus, this finite volume setup does not even correspond to an equilibrium system. Moreover, the global observable $\Omega$ becomes independent of $L_3$ after space-time averaging in this case -- we automatically get the $L_3\to\infty$ limit this way and the two orderings will match. To have an equilibrium system with no $x_4$-dependence, as well as to maintain the volume-dependence of $\Omega$, analogously to $\imu$, we promote $\mu_3$ to be $x_4$-dependent such that $\mubar_3$ is a constant.

Now, making use of the grand canonical approach of Eq.~\eqref{eq:Omegaloc} and setting $\mubar$ and $\mubar_3$ to zero (i.e.\ a globally neutral system with no net currents), we can compute the 
thermodynamic potential.
Interestingly, we need to distinguish the case of $N_e$ being even or odd. For even electric flux, we arrive at 
\begin{equation}
    \Omega(\iE)=\int_0^\infty \frac{\mathrm{d}s}{8\pi^2 s^2}\  e^{-sm^2}  e\iE\coth(e\iE s) \,
    \Theta_3\Biggl[ 0\ ;\ e^{
    -\frac{e\iE L_3^2}{4\tanh(e\iE s)}} \Biggl]\,
    \Theta_4\Biggl[ 0 ;\ e^{
    -\frac{e\iE \beta^2}{4\tanh(e\iE s)}} \Biggl]\,,
\end{equation}
where $\Theta_3$ and $\Theta_4$ are elliptic functions.
In line with Eq.~\eqref{finitevolumesusceptibility}, the susceptibility can be derived by taking the second electric field derivative, and making use of the equality $\iE^2=-E^2$, leading to the result
\begin{equation}
    \left.\frac{\mathrm{d}^2\Omega}{\mathrm{d}(e\iE)^2}\right|_{e\iE=0}=\sum_{j\in\mathds{Z}}\sum_{n\in\mathds{Z}}\int_0^\infty \frac{\mathrm{d}s}{4\pi^2 s^3}\  (-1)^n \,e^{
    -sm^2-\frac{L_3^2j^2}{4s}-\frac{\beta^2 n^2}{4s}} \,\Biggl[  \frac{s^2}{3}-\frac{L_3^2j^2 s}{12} -\frac{\beta^2 n^2 s}{12} \Biggl],
\end{equation}
where we expressed the elliptic $\Theta$ functions via their sum representations.

We can see that the $j\neq0$ contributions are exponentially suppressed. However, to be more careful, we can isolate the $n\neq0$ or $j\neq0$ terms and express them by
\begin{equation}
\begin{split}
    &\sum_{(j,n)\in\mathds{Z}^2\backslash \{0\} }\frac{(-1)^n}{12\pi^2} \Biggl[  K_0\Big(m\sqrt{L_3^2j^2+\beta^2 n^2}\Big)
    -\frac{1}{2} m\sqrt{L_3^2j^2+\beta^2 n^2}\, K_1\Big(m\sqrt{L_3^2j^2+\beta^2 n^2}\Big) \Biggl],
\end{split}
\end{equation}
with modified Bessel functions of the second kind $K_\ell(x)$.
For the Bessel functions, the asymptotic expansion $K_\ell(x)\sim \sqrt{\frac{\pi}{2x}} e^{-x} $ holds in the leading order, confirming the exponential suppression. We can drop all $j\neq0$ terms and the temperature independent $n=0$ contribution, which is ultraviolet divergent (this amounts to carrying out the ultraviolet renormalization mentioned at the end of Sec.~\ref{sec:eqthermo}). The susceptibility is
\begin{equation}
\label{eq:xi1_finitevol}
\begin{split}
    -\underset{L_3\rightarrow\infty}{\lim}\left.\frac{\mathrm{d}^2\Omega}{\mathrm{d}(eE)^2}\right|_{E=0}&=\sum_{n\in\mathds{Z}\backslash\{0\}}\int_0^\infty \frac{\mathrm{d}s}{4\pi^2 s^3}(-1)^n e^{
    -sm^2-\frac{\beta^2 n^2}{4s}}\,\left[ \frac{s^2}{3} -\frac{\beta^2 n^2 s}{12} \right] 
    \\
    &=  -\frac{1}{6\pi^2}\int_m^\infty\frac{\mathrm{d}\omega}{\sqrt{\omega^2-m^2}}\ \Bigl[ 2\nf(\omega)+\omega\frac{\mathrm{d}\nf(\omega)}{\mathrm{d}\omega} \Bigl]\ 
    =\ \xiS\,,
\end{split}
\end{equation}
where $\nf(\omega)=1/(e^{\beta\omega}+1)$ denotes the Fermi-Dirac distribution.
A computation of the odd $N_e$ case differs at the level of $\Omega$, but leads to the same result for the susceptibility.

This concludes the calculation of the bottom rightmost cell of Tab.~\ref{tab:susc}, corresponding to the ordering $\lim_{L_3\to\infty} \int \dd x_3 $ in the grand canonical ensemble. The opposite ordering
$\int \dd x_3 \lim_{L_3\to\infty}$ can be achieved by following the computation of~\cite{Elmfors:1998ee} to take derivatives of the effective (infinite-volume) Lagrangian of~\cite{Elmfors:1994fw}. Here the issue with boundary conditions is absent and one can use real electric fields to begin with. As an alternative, we also consider the analytic continuation of~\eqref{efflag} in the infinite volume limit, where the quantization~\eqref{eq:quantization} disappears and $\iE$ becomes a continuous variable. This leads to
\begin{equation}
\begin{split}
-\mathscr{L}^{T\neq0}_{\text{eff}}(x) = &-\frac{1}{2\pi^{3/2}} \int_{-\infty}^{\infty} 
    \frac{\mathrm{d}\omega }{2\pi } \ff(\omega) \\ &\mathrm{Im}\Biggl(\int_0^\infty\frac{\mathrm{d}s}{s^2}\ e^{-m^2[i+\epsilon]s} \sqrt{eE\coth(eEs)}e^{+i\pi/4}\exp\Biggl[i \frac{\omega^2}{eE\coth(eEs)-i\epsilon} \Biggl]
    \Biggl),
\end{split}
\label{eq:Elmfors}
\end{equation}
where $\ff(\omega)=\theta(\omega)\nf(\omega+\mubar(x))+\theta(-\omega)\nf(-\omega-\mubar(x))$ holds.
In fact, this is a nontrivial check of our approach, and we demonstrate in App.~\ref{app:3} that we obtain the same result as Ref.~\cite{Elmfors:1994fw}. We proceed just as Ref.~\cite{Elmfors:1994fw} does (see also Sec.~3.2 of~\cite{Endrodi:2022wym}): evaluating \eqref{eq:Elmfors} at $\mu=eEx_3$, carrying out the (trivial) integral over $x_3$ and taking the second derivative with respect to $E$, we arrive at the same formula as in the second line of~\eqref{eq:xi1_finitevol}. This shows that the bottom center cell of Tab.~\ref{tab:susc} also equals $\xiS$.

The corresponding computation in the canonical approach also results in $\xiS$, because the additional term $\frac{\partial\Leff}{\partial\mu}(x_3,E,\mu=-eA_0(x_3))$ of Eq.~\eqref{eq:floc} is independent of $x_3$ for homogeneous fields and vanishes in the volume average due to the factor $eA_0(x_3)=-eEx_3$ being odd.

\section{Infinite wavelength and spatial averaging}
\label{sec:4}
We will now consider the ordering of the infinite wavelength limit $k\to 0$ and the spatial average $\int\mathrm{d}x_3$ taken in the computation of the susceptibility. Since we are considering an effective Lagrangian in the infinite volume from the beginning, we can work directly with real electric fields again, as well as real chemical potentials. However, since the effective Lagrangian cannot be obtained for background fields with nonzero $k$ in a closed form, we need to consider a weak-field expansion.

Accounting for the implicit electric field dependence of the chemical potential after setting it to $\mu=- eA_0(x_3)$ results in the formula 
\begin{equation}\label{xigc}
\begin{split}
    -\frac{\mathrm{d}^2\Omega}{\mathrm{d}(eE)^2}\Biggl|_{E=0}=\,-\frac{1}{L_3}\int_{-\frac{L_3}{2}}^\frac{L_3}{2} \mathrm{d}x_3\, \Biggl[
    \frac{\partial^2\Leff(x)}{\partial(eE)^2} &- 
    2\frac{\partial^2\Leff(x)}{\partial(eE)\partial\mu}\frac{\partial eA_0(x)}{\partial(eE)} \\
    &+\frac{\partial^2\Leff(x)}{\partial\mu^2}\left(\frac{\partial eA_0(x)}{\partial(eE)}\right)^2
    \Biggl]_{E=\mu=0}
\end{split}
\end{equation}
in the grand canonical ensemble. In turn, for the canonical ensemble we arrive at (see also Ref.~\cite{Endrodi:2022wym}),
\begin{equation}\label{xic}
    -\frac{\mathrm{d}^2f}{\mathrm{d}(eE)^2}\Biggl|_{E=0}=\,-\frac{1}{L_3}\int_{-\frac{L_3}{2}}^\frac{L_3}{2} \mathrm{d}x_3\, \left[ \frac{\partial^2\Leff(x)}{\partial(eE)^2}-\frac{\partial^2\Leff(x)}{\partial\mu^2}\left(\frac{\partial eA_0(x)}{\partial(eE)}\right)^2\right]_{E=\mu=0}.
\end{equation}
To calculate the ordering $ \lim_{k\to0} \int\dd x_3$, we will first derive the three contributions
\begin{equation}
    \frac{\partial^2\Leff(x)}{\partial(eE)^2}\Biggl|_{E=\mu=0},\qquad
    \frac{\partial^2\Leff(x)}{\partial(eE)\partial\mu}\frac{\partial eA_0(x)}{\partial(eE)}\Biggl|_{E=\mu=0}\quad
    \text{and}\quad
    \frac{\partial^2\Leff(x)}{\partial\mu^2}\left(\frac{\partial eA_0(x)}{\partial(eE)}\right)^2\Biggl|_{E=\mu=0}\,,
    \label{eq:threeterms}
\end{equation}
while retaining the oscillation momentum $k>0$.

In order to be able to represent the three terms in Eq.~\eqref{eq:threeterms} by means of Feynman diagrams, we follow the standard technique to write the effective Lagrangian as the mass-integral of its mass-derivative. This results in
\begin{equation}\label{eq:Feynman}
\begin{split}
     -\frac{\partial^2 \mathscr{L}_{\text{eff}}(x)}{\partial (eE)^2}\Biggl|_{E=\mu=0}
    &= 
    \int\dd m
\begin{tikzpicture}[baseline=0.0cm]
    \filldraw [black] (0,0) circle (1.5pt);
    \begin{feynman}[small]
        \draw node at ($(-0.2cm,0.2cm)$) {$x$};
        \vertex at ($(0cm,0cm)$) (x);
        \vertex at ($(0.425cm,0.425cm)$) (y);
        \vertex at ($(0.425cm,-0.425cm)$) (z);
        \vertex at ($(0.425cm,0.85cm)$) (t1);
        \vertex at ($(0.425cm,-0.85cm)$) (t2);
    \diagram*{
        (x) --[fermion, quarter left] (y),
        (z) --[fermion, quarter left] (x),
        (y) --[fermion, half left] (z),
        (y) --[photon] (t1),
        (z) --[photon] (t2),
    };
    \end{feynman}
\end{tikzpicture}\Biggl|_{E=\mu=0}\\
-\frac{\partial^2\Leff(x)}{\partial(eE)\partial\mu}\frac{\partial eA_0(x)}{\partial(eE)}\Biggl|_{E=\mu=0} &=
\int\mathrm{d} m\ \frac{\partial^2}{\partial(eE)\partial\mu}
\begin{tikzpicture}[baseline=0.0cm]
\filldraw [black] (0,0) circle (1.5pt);
\begin{feynman}[small]
\draw node at ($(-0.2cm,0.2cm)$) {$x$};
\vertex at ($(0cm,0cm)$) (x);
\vertex at ($(0.85cm,0cm)$) (y);
\vertex at ($(y) + (0.55cm,0cm)$) (t);
\diagram*{
(x) --[fermion, half left] (y),
(y) --[fermion, half left] (x),
(y) --[photon, edge label= \(A_\mu\)] (t)
};
\end{feynman}
\end{tikzpicture} \frac{\partial eA_0(x)}{\partial(eE)} \Biggl|_{E=\mu=0}\\
-\frac{\partial^2\Leff(x)}{\partial\mu^2}\left(\frac{\partial eA_0(x)}{\partial(eE)}\right)^2\Biggl|_{E=\mu=0} &=
\int\mathrm{d}m\ \frac{\partial^2}{\partial\mu^2} 
\begin{tikzpicture}[baseline=0.0cm]
\filldraw [black] (0,0) circle (1.5pt);
\begin{feynman}[small]
\draw node at ($(-0.2cm,0.2cm)$) {$x$};
\vertex at ($(0cm,0cm)$) (x);
\vertex at ($(0.85cm,0cm)$) (y);
\diagram*{
(x) --[fermion, half left] (y),
(y) --[half left] (x)
};
\end{feynman}
\end{tikzpicture}\  \left(\frac{\partial eA_0(x)}{\partial(eE)}\right)^2 \Biggl|_{E=\mu=0}
\end{split}
\end{equation}
where the constant of the mass integration is chosen such that the terms vanish for $m\to\infty$.
The Feynman diagrams above include a scalar insertion at $x$ (resulting from the $m$-derivative) and photon legs with factors $eA_0$ attached to them (due to the electric field derivative). Notice that these are mixed representation diagrams in the sense that they are parameterized by the position of the scalar insertion but the momentum flowing in and out on the photon legs due to the oscillation $k$ of the background field.

We solve the three sum-integrals of Eq~(\ref{eq:Feynman}) by first solving the Matsubara sum via a contour integration and identifying the finite temperature contributions, simplifying the angular integration and at last performing the mass integration. For further details, see App.~\ref{app:4}. This yields the integral representations over the momentum variable 
\begin{equation}\label{eq:2legsuscont}
\begin{split}
    -\frac{\partial^2 \mathscr{L}_{\text{eff}}(x)}{\partial (eE)^2}&\Biggl|_{E=\mu=0}
     = \,\int_0^\infty\dd p \, 
    \frac{p\, n_f(E_p)}{8 \pi ^2 k^5
   E_p }\, \Bigg[
     k^2 \left(4 E_p^2-k^2\right) \log \left(\frac{(k+2 p)^2}{(k-2 p)^2}\right)+8 k^3 p
    - \cos \Big(2 k x_3\Big)\\ &\Bigg(4 k^3 p
    +4 \left(k^2-p^2\right) E_p^2 \log \left(\frac{(k+p)^2}{(k-p)^2}\right)-\frac{1}{2} \left(k^2-4 p^2\right) \Big(k^2+4 E_p^2 \log \left(\frac{(k+2 p)^2}{(k-2 p)^2}\right) \Big)\Bigg)
    \Bigg],
\end{split}
\end{equation}
\begin{equation}\label{eq:1legsuscont}
    \frac{\dd^2 \mathscr{L}_{\text{eff}}(x)}{\partial (eE) \partial\mu}\frac{\partial A_0(x)}{\partial (eE)}\Biggl|_{E=\mu=0} = \,\int_0^\infty \dd p\,
    \sin^2 (k x_3)
    \frac{p\,  n_f(E_p)}{8 \pi ^2 k^3  E_p} \,
    \Biggl[\left(8 m^2+12 p^2 -k^2 \right) \log \left(\frac{(k-2 p)^2}{(k+2 p)^2}\right)-8 k p\Biggl],
\end{equation}
and, finally 
\begin{equation}\label{eq:0legsuscont}
    -\frac{\partial^2\Leff(x)}{\partial\mu^2}\left(\frac{\partial eA_0(x)}{\partial(eE)}\right)^2\Biggl|_{E=\mu=0} = \,\int_0^\infty \dd p\, \sin^2 (k x_3)\,
    \frac{2 \left(m^2+2 p^2\right) n_f(E_p) }{\pi ^2 k^2 E_p }.
\end{equation}

Taking the volume average 
$\int\dd x_3$ of Eqs.~(\ref{eq:1legsuscont}) and (\ref{eq:0legsuscont}) in the limit $L_3\to\infty$ and at finite $k$
replaces $\sin^2(kx_3)$ by its average $1/2$, while in Eq.~(\ref{eq:2legsuscont}) the term proportional to $\cos(2kx_3)$ fully vanishes.\footnote{Note that this also holds in a finite volume if the oscillation wavelength and the spatial extent are commensurate, i.e.\ $k=2\pi n/L_3$ with $n\in\mathds{Z}^+\backslash\{0\}$. 
}
After 
this limit is taken, we form an expansion with respect to $k$ and obtain
\begin{equation}\label{eq:2ndDiagramkint0}
    -\lim_{L_3\to\infty}\, \frac{1}{L_3}\int_{-\frac{L_3}{2}}^{\frac{L_3}{2}}\dd x_3\,  \frac{\partial^2 \mathscr{L}_{\text{eff}}(x)}{\partial (eE)^2}\Biggl|_{E=\mu=0} =\,
    \frac{1 }{2k^2} \chimumu
    +\frac{1}{2}\xiW +\mathcal{O}(k^2),
\end{equation}
\begin{equation}\label{eq:2ndDiagramkint}
    -\lim_{L_3\to\infty}\,\frac{1}{L_3}\int_{-\frac{L_3}{2}}^{\frac{L_3}{2}}\dd x_3\,\frac{\dd^2 \mathscr{L}_{\text{eff}}(x)}{\partial (eE) \partial\mu}\frac{\partial A_0(x)}{\partial (eE)}\Biggl|_{E=\mu=0} =\, \frac{1}{2k^2}\chimumu  + \frac{1}{4}(\xiW - \xiS) +\mathcal{O}\left(k^2\right)
\end{equation}
and
\begin{equation}
    -\lim_{L_3\to\infty}\,\frac{1}{L_3}\int_{-\frac{L_3}{2}}^{\frac{L_3}{2}}\,\dd x_3
    \frac{\partial^2\Leff(x)}{\partial\mu^2}\left(\frac{\partial eA_0(x)}{\partial(eE)}\right)^2\Biggl|_{E=\mu=0} =\,
    \frac{1 }{2k^2} \chimumu,
\end{equation}
where $\chimumu =- \partial^2_\mu\,\Omega$ is the fermion charge susceptibility with respect to the chemical potential $\mu$. 
Consulting Eqs.~(\ref{xigc}) and (\ref{xic}), the susceptibilities in the present ordering can be extracted. In particular, we observe that terms containing $\chimumu$, being IR divergent in the limit $k\to0$, cancel both in $\Omega$ and in $f$. These are precisely the terms that quantify the response of the inhomogeneous charge profile, see Eq.~\eqref{eq:omegafree}. In turn, the IR finite terms combine in different ways, producing $\xiW$ in the canonical and $\xiS$ in the grand canonical ensemble. We highlight that Eq.~(\ref{eq:2ndDiagramkint}) is the term responsible for this mismatch. We remind the reader that factors of $1/2$ arise due to the normalization of the oscillatory vector potentials, cf.\ footnote~\ref{fn:factor2}.

Turning to the opposite ordering and taking the $k\to0$ limit of Eq.~(\ref{eq:2legsuscont}) yields
\begin{equation}\label{eq:1contk0}
    -\lim_{k\to0}\,
    \frac{\partial^2\mathscr{L}_{\text{eff}}(x)}{\partial (eE)^2}\Biggl|_{E=\mu=0} =\, \chimumu x_3^2 + \xiS ,
\end{equation}
now with a finite contribution originating from the $\cos(2kx_3)$ term of Eq.~(\ref{eq:2legsuscont}), resulting in the appearance of $\xiS$ instead of $\xiW$ as in Eq.~(\ref{eq:2ndDiagramkint0}).
Performing the same limit for Eq.~(\ref{eq:1legsuscont}) and (\ref{eq:0legsuscont}) creates the purely position-dependent terms
\begin{equation}\label{eq:2contk0}
    -\lim_{k\to0}\, \frac{\partial^2 \mathscr{L}_{\text{eff}}(x)}{\partial (eE) \partial\mu}\frac{\partial A_0(x)}{\partial (eE)}\Biggl|_{E=\mu=0} =\,
    \chimumu x_3^2
\end{equation}
and
\begin{equation}\label{eq:3contk0}
    -\lim_{k\to0}\,
    \frac{\partial^2\Leff(x)}{\partial\mu^2}\left(\frac{\partial eA_0(x)}{\partial(eE)}\right)^2\Biggl|_{E=\mu=0} =\,
    \chimumu x_3^2.
\end{equation}
Consulting Eq.~(\ref{xigc}) and (\ref{xic}) again, the IR divergent terms again cancel, but this time the susceptibility yields $\xiS$ both for the canonical and the grand canonical ensemble. Eqs.~(\ref{eq:1contk0}), (\ref{eq:2contk0}) and (\ref{eq:3contk0}) are consistent with the effective Lagrangian $\Leff$ of \cite{Elmfors:1994fw,Elmfors:1998ee}.

\section{Hadron resonance gas model}
As an application of our findings, next we study the electric susceptibility $\xi$ in a simple hadron resonance gas-type model, which is relevant for the description of the low-temperature region of QCD. For simplicity, we only consider the lightest hadrons i.e.\ pions in the model.
This means that instead of fermion propagators, we need to work with scalar fields.

To calculate $\xi$ within the model, we constrain the discussion to the canonical ensemble and the ordering $\lim_{k\to0} \int\dd x_3$, i.e.\ the one that results in $\xiW$ for fermions.
This choice enables a direct comparison to lattice QCD simulations. This is because on the lattice, this particular setup can be implemented conveniently by means of a two-point function at a fixed mass; see Eq.~(\ref{eq:scalarchixi}) below. Hence, the lattice measurement of the so defined susceptibility is reduced to determining a current-current correlator,
see Ref.~\cite{Endrodi:2023wwf} for a more detailed discussion.

Besides $\xi$, we also consider the magnetic susceptibility $\chi$, defined as the leading response of the thermodynamic potential with respect to weak background magnetic fields $B$. Thus, the magnetic equivalent of~\eqref{eq:defxi} is the definition $\chi=-\partial^2 \Omega/\partial (eB)^2$ evaluated at $B=0$. For $\chi$, no infrared ambiguities arise, since magnetic fields do not lead to a singular rearrangement of charge densities. Therefore, the canonical and grand canonical ensembles are equivalent, and the spatial averaging and infinite wavelength limits commute with each other in this case.

In Ref.~\cite{Endrodi:2022wym}, it was shown how to relate the response at wavelength $2\pi/k$ to the photon polarization diagram with momentum flow $k$. In particular, the susceptibilities for spin-zero bosonic particles can be expressed as, see also~\cite{Buividovich:2021fsa}
\begin{equation}
    \xi = \frac{1}{2e^2}\lim_{k\to0}\lim_{k^0\to0} \frac{\mathrm{d}^2 \Pi_{00}^{T\neq 0}(k^0,k) }{\mathrm{d}k^2},\qquad \chi = \frac{1}{4e^2}\lim_{k\to0}\lim_{k^0\to0} \frac{\mathrm{d}^2 \Pi_{jj}^{T\neq 0}(k^0,k) }{\mathrm{d}k^2}\,,
    \label{eq:scalarchixi}
\end{equation}
using an implied sum over the indices $j$ and the finite temperature contribution to the photon polarization diagram in scalar QED,
\begin{equation}
 \label{eq:Pimunudef}
 \Pi_{\mu\nu}=
 \begin{tikzpicture}[baseline=0.0]
\begin{feynman}[small]
\vertex at ($(0,0)$) (x);
\vertex at ($(0.85cm,0)$) (y);
\vertex at ($(x)-(0.85cm,0)$) (p);
\vertex at ($(y)+(0.85cm,0)$) (t);
\diagram*{
(x) --[fermion, half left, momentum = \(p+k\)] (y);
(y) --[fermion, half left, momentum = \(p\)] (x);
(p) --[photon, momentum = \(k\), edge label'=\(\mu\)] (x);
(y) --[photon,momentum = \(k\), edge label'=\(\nu\)] (t);
};
\end{feynman}
\end{tikzpicture}
    = (-ie)^2\int \frac{\mathrm{d}^4 p}{(2\pi)^4}(k+2p)^\mu (k+2p)^\nu \Sb(p) \Sb(k+p)\,,
\end{equation}
where the solid lines denote scalar propagators. The susceptibilities in Eq.~\eqref{eq:scalarchixi} are already renormalized, which is why the vacuum contribution $\Pi_{\mu\nu}^{T=0}$ is absent.

Above we used the Minkowski metric $g_{\mu\nu}=\textmd{diag}(1,-1,-1,-1)$.
The evaluation of this diagram can be performed by following the steps outlined in \cite{Endrodi:2022wym}. 
The details are included in App.~\ref{app:5}.
The final result for the electric susceptibility is
\begin{equation}
    \xi=  \frac{1}{12\pi^2}\int_m^\infty\frac{\mathrm{d}\omega}{\sqrt{\omega^2-m^2}}\Big( \nb(\omega) +\omega\, n_B'(\omega) \Big),
    \label{eq:resBxi}
\end{equation}
while for the magnetic susceptibility we obtain
\begin{equation}
    \chi = -\frac{1}{12\pi^2}\int_m^\infty \frac{\mathrm{d}\omega}{\sqrt{\omega^2-m^2}}\,\nb(\omega) = -\frac{1}{48\pi^2}\int_0^\infty\frac{\mathrm{d}t}{t}\ e^{-m^2t/T^2}\left(
    \Theta_3\left[ 0\ ;\ e^{-1/(4t)} \right]
    -1\right)\,.
    \label{eq:resBchi}
\end{equation}
Here, $\nb(\omega)=1/(e^{\beta\omega}-1)$ is the Bose-Einstein distribution.
The result for $\chi$ agrees with the elliptic $\Theta_3$ function representation of Ref.~\cite{Bali:2020bcn}.

\begin{figure}
\center
\includegraphics[width=0.6\textwidth]{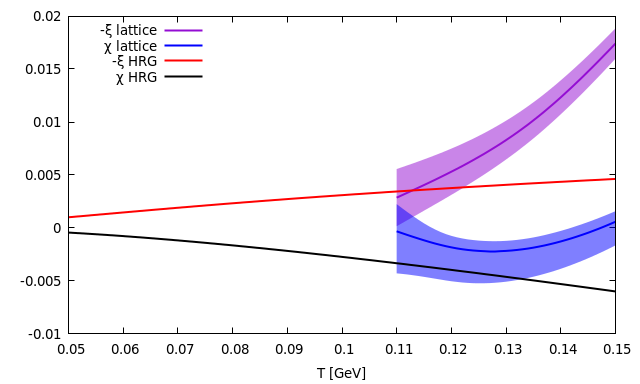}
\caption{The negative of the electric susceptibility $\xi$ and the magnetic susceptibility $\chi$ of a simple hadron resonance gas model, consisting only of the pions compared to continuum extrapolated lattice results~\cite{Endrodi:2023wwf}.}
\label{fig:hadrongasusc}
\end{figure}

To arrive at a quantitative prediction relevant for the low-temperature region of QCD, we use Eqs.~\eqref{eq:resBxi} and~\eqref{eq:resBchi} with $m=m_\pi=135\textmd{ MeV}$ for the charged pions. The so obtained results are plotted in Fig.~\ref{fig:hadrongasusc} and compared to the continuum extrapolated lattice QCD results of Ref.~\cite{Endrodi:2023wwf}. The data and the model are found to match well at temperatures up to about $T\approx120\textmd{ MeV}$.
We note that the magnetic susceptibility is found to be negative, signaling the diamagnetism of the medium consisting of thermal pions. This is due to the spinless nature of pions, which couple to the magnetic field only via angular momentum, giving a diamagnetic contribution due to Lenz's law. Regarding the electric susceptibility, we find that $\xi<0$ and decreases as the temperature grows. This signals that charges taking part in thermal fluctuations generate a polarization which opposes the background electric field.

\section{Conclusion}
\label{sec:conclusions}

In this paper we discussed the impact of background electric fields on a charged plasma at high temperature using leading-order perturbation theory. This setup is relevant both for the high-temperature behavior of QCD and QED plasmas, as well as for the modeling of the low-energy region of strongly interacting matter within hadron resonance gas approaches.

In general, at nonzero temperature electric fields induce an inhomogeneous equilibrium charge density profile that is singular for homogeneous backgrounds in the infinite volume. Theoretically this profile extends over macroscopic length scales and describes a rearrangement of the plasma's charged particles.
Therefore, we isolate the response of the singular charge distribution to quantify the impact of the medium on microscopic scales, induced by the quantum behavior of the system.
This isolation leads to ambiguities in the definition of the equation of state in the presence of the electric field. 
As a prime example, the response for weak fields characterized by the electric susceptibility $\xi$ is affected by how the infrared regularization to perform the isolation is implemented and eventually removed. Building on the seminal results by Weldon~\cite{Weldon:1982aq} and the finite-temperature calculations~\cite{Loewe:1991mn,Elmfors:1994fw,Elmfors:1998ee,Gies:1998vt,Gies:1999xn} based on the Schwinger propagator~\cite{Schwinger:1951nm}, the mismatch for the susceptibility, $\xiS\neq\xiW$, was pointed out in~\cite{Endrodi:2022wym}.

Here we carried out a comprehensive analysis of this problem and identified the reason for the mismatch: infrared regularizations involving, on the one hand, a homogeneous field in finite system size $L_3$ and, on the other hand, an oscillatory field with finite wavelength $\sim1/k$ represent different physical realizations in the canonical statistical ensemble. In particular, the spatial averaging over the entire system does not commute with removing the infrared regularization in the latter case. In addition, we also demonstrated that for oscillatory background fields, the grand canonical and canonical treatments of this system differ even in the $k\to0$ limit. 
In turn, for finite volumes, such ambiguities were not found, see the summary in Tab.~\ref{tab:susc}.

\begin{figure}
 \centering
\includegraphics[width=0.6\textwidth]{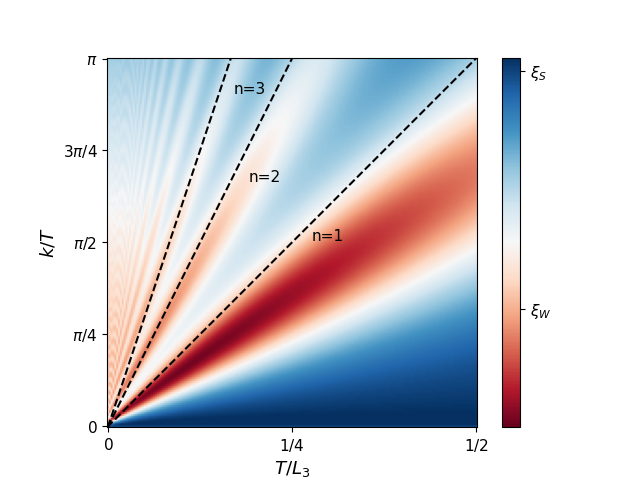}
\caption{\label{fig:illustr} 
The IR regulated electric susceptibility in the canonical ensemble -- obtained via the difference of Eqs.~(\ref{eq:2legsuscont}) and~(\ref{eq:0legsuscont}), along with the normalization of Eq.~(\ref{eq:normalization}) --  for nonzero momentum $k$ averaged over a finite volume $L_3$ for $m/T=1$. The point in the bottom left corresponds to the limit of a homogeneous electric field in the infinite volume. The result for the susceptibility depends on the angle used to approach this limiting point. 
}
\end{figure}

To illustrate the inequivalence of electric susceptibilities obtained in the limits of removing the infrared regularizations, $L_3\to\infty$ and $k\to0$, 
it is instructive to interpolate the result to intermediate values of $L_3$ and $k$. This 
is provided in Fig.~\ref{fig:illustr}, where the susceptibility~\eqref{xic} of the canonical approach is plotted for nonzero $k$ and finite volume average over size $L_3$. To enable a consistent comparison of these setups, we include the normalization factor corresponding to the average vacuum energy of the electric field, cf.\ footnote~\ref{fn:factor2},
\begin{equation}\label{eq:normalization}
    \frac{1}{L_3}\int_{-\frac{L_3}{2}}^{\frac{L_3}{2}}\dd x_3\, \cos^2(kx_3)
    =\begin{cases}
    \ 1; & k\to0 \\
    \ \frac{1}{2}; &  L_3\to\infty 
    \end{cases}.
\end{equation}
The target system, involving homogeneous electric fields extending over an infinite volume (without the contribution of the macroscopic charge distribution), corresponds to the bottom left point. This point can be approached via homogeneous fields subject to a finite volume average $L_3$ (the horizontal axis subject to $k=0$), or via oscillatory fields with momentum $k$ in the infinite volume (vertical axis constrained by $1/L_3=0$).
These paths result in the mismatching susceptibilities $\xiS$ and $\xiW$.
Further, the trajectories subject to the momentum quantization $k=2\pi n/L_3$ with $n\in\mathds{Z}^+$ also result in $\xiW$.
Overall, a non-analytic structure appears, ultimately leading to the non-commutativity of the limits involved.

Finally, we constructed a hadron resonance gas-type model for the low-temperature QCD medium. For this, we used the Weldon-type definition involving oscillatory fields -- for which lattice QCD simulations are available -- and considered only the leading contribution due to charged pions. The results are found to agree remarkably with the those obtained by first-principles lattice QCD simulations, both for the electric as well as for the magnetic susceptibility, providing a highly non-trivial check of our analytical approach.

\acknowledgments

G.~E. and G.~M. are grateful for insightful discussions with Zolt\'an R\'acz.
L.~S.\ has been supported by the Research Council of Finland projects 353772 and 354533.
Moreover, this work was funded by the DFG (Collaborative Research Center CRC-TR 211 ``Strong-interaction matter under
extreme conditions'' - project number 315477589 - TRR 211), the Hungarian National Research, Development and Innovation Office - NKFIH (Research Grant Hungary 150241) and the European Research Council (Consolidator Grant 101125637 CoStaMM). Views and opinions expressed are however those of the authors only and do not necessarily reflect those of the European Union or the European Research Council. Neither the European Union nor the granting authority can be held responsible for them.

\appendix
\section{Finite volume propagator}
\label{app:1}

In this appendix, we present the fermion propagator in the presence of a homogeneous imaginary electric field background field and limited to a finite spatial extent $L_3$ and temperature $T$.
Here we work with the imaginary electric field $\iE$ and imaginary chemical potential $\imu$. The imaginary time is denoted by $x_4$.

For the propagator $G$ (the inverse of the Dirac operator), the boundary conditions~\eqref{eq:bc}  result in the constraints,
\begin{equation}
\begin{split} 
    G(x+L_3\hat{e}_3,y)&=e^{-i\mu_3L_3+ie\iE L_3x_4}\ G(x,y)\\
    G(x,y+L_3\hat{e}_3)&= G(x,y) \ e^{i\mu_3L_3-ie\iE L_3y_4}\\
    G(x+\beta\hat{e}_4,y)&= -e^{i \imu\beta} \ G(x,y)\\
    G(x,y+\beta\hat{e}_4)&= - G(x,y) \ e^{-i \imu\beta},
\end{split}
\end{equation}

For our purposes, the image sum construction of Ref.~\cite{Adhikari:2023fdl} can be written as
\begin{equation}
    G(x,y) = \sum_{n_3,n_4\in\mathds{Z}} e^{+i\mu_3L_3n_3-ie\iE L_3x_4n_3} (-1)^{n_4}e^{-i \imu\beta n_4} \,G_{\infty}(x +L_3n_3\hat{x}_3+\beta n_4\hat{x}_4 ,y),
\end{equation}
where $G_{\infty}$ is the infinite volume propagator~\cite{Schwinger:1951nm}, continued to Euclidean time and imaginary electric field.

Performing the infinite sums, the finite volume propagator reads
\begin{equation}\label{fullGfunction}
    G(x,y)=(\slashed{p}_x+m)\,     \text{diag}[G_+(x,y),G_-(x,y),G_-(x,y),G_+(x,y)],
\end{equation}
where the differentiation involved in the covariant momentum operator $\slashed{p}_x$ acts on the coordinate $x$ and the Weyl representation of the Dirac matrices $\gamma^\mu$ was used.\footnote{For the $\gamma$-matrices the real time relation $\{\gamma^\mu,\gamma^\nu\}=2\eta^{\mu\nu}$ holds. For the contraction $\slashed{p}$ this implies, by the relation $x_4=-ix_0$ that the term $\gamma^0iD_0=\gamma^0i(-iD_4)$ is contained.}
The functions in the second factor of~\eqref{fullGfunction} are given by scalar propagators with masses of the form $m_{\pm}^2=m^2\pm e\iE$, which can be represented through the Schwinger parametrization,
\begin{equation}
    G_\pm(x,y)
    = \bra{x}\left[-p^2+m_\pm^2\right]^{-1}\ket{y}
    =\int_0^\infty \mathrm{d}s\ e^{-sm_{\pm}^2}
    g^\parallel_{L_3,\beta}(x_\parallel,y_\parallel|s) g^\bot(x_\bot,y_\bot|s),
\end{equation}
to decompose the function into the coordinates $x_\bot=(x_1,x_2)$, which couple to the background field and $x_\parallel=(x_3,x_4)$, which decouple from the field, just like it was done in Ref.~\cite{Adhikari:2023fdl}.

In the perpendicular direction $x_\bot$ the propagator reads
\begin{equation}\label{gfree}
    g^\bot(x_\bot,y_\bot|s)=
    \frac{1}{4\pi s}\,e^{-(x_\bot-y_\bot)^2/(4s)}\,.
\end{equation}
For the parallel component $x_\parallel$, we need to distinguish between two cases. For even values of the imaginary electric flux quantum $N_e$, we obtain
\begin{equation}
\begin{split}\label{geven}
    g^{\parallel\ N_e \;\text{even}}_{L_3,\beta}(x_\parallel,y_\parallel|s) = 
    &\frac{e\iE}{4\pi\sinh(e\iE s)} \exp\Biggl[ 
    -\frac{e\iE}{4\tanh(e\iE s)}[x_\parallel-y_\parallel]^2  + i\frac{e\iE}{2}[x_4-y_4][x_3+y_3] 
    \Biggl] \\ 
    &\Theta_3\Biggl[ i\frac{e\iE L_3}{4\tanh(e\iE s)}[x_3-y_3]  + \frac{L_3\mu_3}{2}  -\frac{e\iE L_3}{4}[x_4+y_4]\ ; \ e^{ -\frac{e\iE L_3^2}{4\tanh(e\iE s)} }\Biggl] \\ 
    &\Theta_4\Biggl[ i\frac{e\iE \beta}{4\tanh(e\iE s)}[x_4-y_4] -  \frac{\beta\imu}{2} + \frac{e\iE\beta}{4}[x_3+y_3] \ ; \ e^{ -\frac{e\iE\beta^2}{4\tanh(e\iE s)} } \Biggl]\,,
\end{split}
\end{equation}
while for odd values of $N_e$, we arrive at
\begin{equation}
\begin{split}\label{godd}
    g^{\parallel\ N_e \;\text{odd}}_{L_3,\beta}(x_\parallel,y_\parallel|s) = &\frac{e\iE}{4\pi\sinh(e\iE s)} \exp\Biggl[ 
    -\frac{e\iE}{4\tanh(e\iE s)}[x_\parallel-y_\parallel]^2 + i\frac{e\iE }{2}[x_4-y_4][x_3+y_3] 
    \Biggl] \\
    \Biggl( &\Theta_2\Biggl[ i\frac{e\iE L_3}{2\tanh(e\iE s)}[x_3-y_3]  + {L_3\mu_3} -\frac{e\iE L_3}{2}[x_4+y_4]\ ; \ e^{ -\frac{e\iE L_3^2}{\tanh(e\iE s)} } \Biggl] \\
    &\Theta_3 \Biggl[ i\frac{e\iE \beta}{4\tanh(e\iE s)}[x_4-y_4]  -  \frac{\beta\imu}{2} + \frac{e\iE\beta}{4}[x_3+y_3] \ ; \ e^{ -\frac{e\iE\beta^2}{4\tanh(e\iE s)} } \Biggl] \\
    + &\Theta_3\Biggl[ i\frac{e\iE L_3}{2\tanh(e\iE s)}[x_3-y_3]  + {L_3\mu_3} -\frac{e\iE L_3}{2}[x_4+y_4]\ ; \ e^{ -\frac{e\iE L_3^2}{\tanh(e\iE s)} } \Biggl]  \\ 
    &\Theta_4 \Biggl[ i\frac{e\iE \beta}{4\tanh(e\iE s)}[x_4-y_4]  -  \frac{\beta\imu}{2} + \frac{e\iE\beta}{4}[x_3+y_3] \ ; \ e^{ -\frac{e\iE\beta^2}{4\tanh(e\iE s)} } \Biggl] \  \Biggl), 
\end{split}
\end{equation}
represented by elliptic $\Theta$ functions.
These functions enter the effective Lagrangian~\eqref{efflag} of the main text.

\section{Charge density and currents}
\label{app:2}

In this appendix we calculate the local charge density and the local electric current in the direction of the electric field using the propagator derived in App.~\ref{app:1}. This will help us in order to explore the nature of the system in equilibrium.
The density and the current~\eqref{eq:densitycurrent} are defined by
\begin{equation}
    \langle\bar\psi(x)\gamma^0\psi(x)\rangle = - \mathrm{Tr}[G(x,x)\gamma^0], \qquad \langle\bar\psi(x)\gamma^3\psi(x)\rangle =
    -\mathrm{Tr}[G(x,x)\gamma^3]\,.
\end{equation}
Simplifying the traces and covariant derivatives, we find 
\begin{equation}\label{chargedensityeq}
    \langle\bar\psi(x)\gamma^0\psi(x)\rangle = -
    i\int_0^\infty\frac{\mathrm{d}s}{2\pi s}\ e^{-m^2s} \frac{qE}{\sinh(qEs)} \frac{\mathrm{d}}{\mathrm{d}\imu}
    g^\parallel_{L_3,\beta}(x_\parallel,x_\parallel|s)
\end{equation}
and
\begin{equation}\label{eq:current}
    \langle\bar\psi(x)\gamma^3\psi(x)\rangle = 
    -i\int_0^\infty\frac{\mathrm{d}s}{2\pi s}\ e^{-m^2s} \frac{qE}{\sinh(qEs)} \frac{\mathrm{d}}{\mathrm{d}\mu_3}
    g^\parallel_{L_3,\beta}(x_\parallel,x_\parallel|s),
\end{equation}
where $g^\parallel_{L_3,\beta}(x_\parallel,x_\parallel|s)$ is given by Eq.~\eqref{geven} or~\eqref{godd}, depending on whether the flux quantum is even or odd.

\begin{figure}
\includegraphics[width=0.5\textwidth]{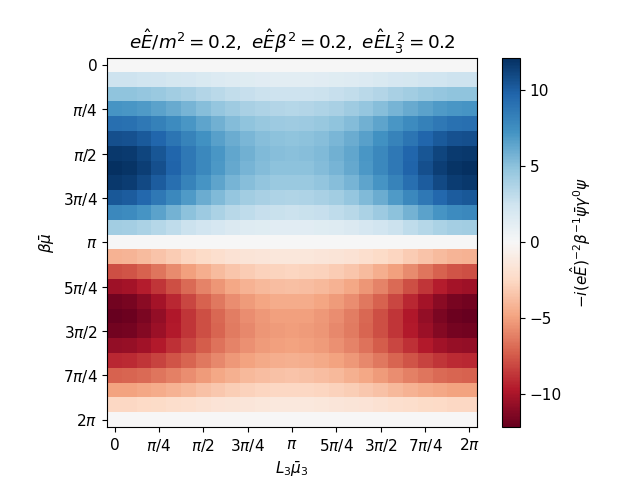}
\includegraphics[width=0.5\textwidth]{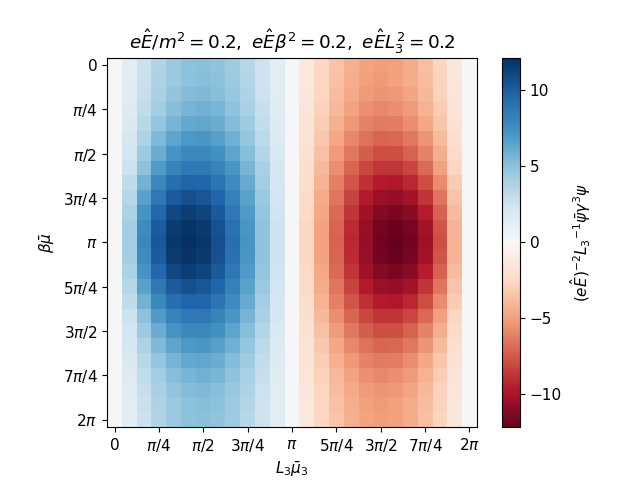}
\caption{A visualized charge distribution $\langle\bar\psi(x)\gamma^0\psi(x)\rangle$ on the left and a visualized charge current $\langle\bar\psi(x)\gamma^3\psi(x)\rangle$ on the right as a function of the local chemical potentials $\mubar$ and $\mubar_3$ for even $N_e$.}\label{fig:currentplot}
\end{figure}

Notice 
that the dependence of $g^\parallel_{L_3,\beta}(x_\parallel,x_\parallel|s)$ on imaginary time and space appears purely through the local chemical potentials $\mubar(x)=\imu-e\iE x_3$ and $\mubar_3(x)=\mu_3-e\iE x_4$. Moreover, it is apparent that both observables~\eqref{chargedensityeq} and~\eqref{eq:current} depend on the imaginary time -- despite that in equilibrium, the dependence on $x_4$ needs to drop out in equal time correlations, see Fig.~\ref{fig:currentplot} for a visualization of the charge distribution and charge current.
Only when the infinite volume limit is taken, does the dependence on $x_4$ vanish and the equilibrium is
reached.
This problem is resolved in the finite volume by the construction discussed below Eq.~\eqref{eq:densitycurrent}. Thus, we promote $\mu_3$ to be $x_4$-dependent so that $\mubar_3$ is a constant. 
Since we would like to describe a system without a charge current present after the infinite volume limit, this constant is chosen to be zero. This implies that the charge density becomes independent of $x_4$ and the current vanishes altogether. 

\section{Crosscheck of the effective Lagrangian}
\label{app:3}

In this appendix, we compare our analytical findings to other existing results in the literature.
In particular, the mass derivative of the effective Lagrangian (i.e.\ the chiral condensate) has already been derived for a finite volume system in the past in Ref.~\cite{PhysRevD.108.076002}.  

The expression found for the chiral condensate in Ref.~\cite{PhysRevD.108.076002} can be transformed into
\begin{equation}\label{eq:correa}
\begin{split}
    \langle\bar\psi\psi\rangle  
    = -m\int_0^\infty \frac{\mathrm{d}s}{4\pi^2 s}\ & e^{-sm^2}  q\iE\coth{q\iE s} \,
    \Theta_3\Biggl[ \frac{L_3\mu_3}{2}\ ;\ e^{ -\frac{e\iE L_3^2}{4\tanh(e\iE s)}} \Biggl]  
    \Theta_4\Biggl[ -\frac{\beta\imu}{2}\ ;\ e^{ -\frac{e\iE\beta^2}{4\tanh(e\iE s)}} \Biggl].
\end{split}
\end{equation}
We find our result to disagree with the result of Ref.~\cite{PhysRevD.108.076002} in general. We only agree under the specific case when $N_e$ is even and $\mubar(x)$ and $\mubar_3(x)$ are both replaced by constants.

To highlight this specific case, we compare the global expression~\eqref{eq:correa} to the formulae derived here. In our formalism, the local chiral condensate is
\begin{equation}\label{localqcond}
    \langle \bar\psi(x) \psi(x) \rangle = -m\int_0^\infty \frac{\mathrm{d}s}{\pi s}\  e^{-sm^2}  \cosh(qEs)  g^\parallel_{L_3,\beta}(x_\parallel,x_\parallel|s),
\end{equation}
and the propagator factor $g^\parallel_{L_3,\beta}$ differs whether the electric field quantum $N_e$ is even or odd, cf.\ Eqs.~\eqref{geven} and~\eqref{godd}.

Performing the replacement $\mubar(x)=\imu-e\iE x_3\to\imu$ and $\mubar_3(x)=\mu_3-e\iE x_4\to\mu_3$, rendering the condensate $x$-independent and averaging over space-time, we only find agreement in the even $N_e$ case, but not for the odd $N_e$ case. This is because only in the even $N_e$ case does $g^\parallel_{L_3,\beta}$ reduce to a simple product of two $\Theta$ functions, see Eq.~(\ref{geven}) and (\ref{godd}). Furthermore, would we average over space-time without a constant $\mubar(x)$ and $\mubar_3(x)$ replacement, we would end up with
\begin{equation}\label{qcond}
    \langle\bar\psi\psi\rangle = i\int \frac{\mathrm{d}^4x}{\beta V}\ \langle \bar\psi(x) \psi(x) \rangle = -m\int_0^\infty \frac{\mathrm{d}s}{4\pi^2 s}\  e^{-sm^2}  e\iE  \coth(e\iE s),
\end{equation}
i.e.\ the condensate in the vacuum and in infinite volume, $L_3=\beta=\infty$.

In order to understand why we only find agreement with Ref.~\cite{PhysRevD.108.076002} when the replacements $\mubar(x)=\imu-e\iE x_3\to\imu$ and $\mubar_3(x)=\mu_3-e\iE x_4\to\mu_3$ are made, we note that the derivation in Ref.~\cite{PhysRevD.108.076002} did not account for twisted boundary conditions, but was based on periodic boundary conditions. The disagreement in the odd $N_e$ case occurs, since the electric field quantization is not accounted for in Ref.~\cite{PhysRevD.108.076002}.

As a further consistency check we show next that our findings for the chiral condensate in the infinite volume are consistent with the findings of Ref.~\cite{Elmfors:1994fw} after an analytic continuation to real electric fields $E$ and real chemical potentials $\mu$ is performed. Such an analytic continuation is only possible after the infinite volume limit has been taken, since in the finite volume the imaginary electric field is quantized.

Taking the infinite volume limit of our local expression, we find
\begin{equation}\label{eq:infvolchiral}
\begin{split}
    \langle \bar\psi(x) \psi(x) \rangle  
    &= -m\int_0^\infty \frac{\mathrm{d}s}{4\pi^2 s}\  e^{-sm^2}  q\iE\coth{q\iE s} \ 
    \Theta_4\Biggl[ -\frac{\beta\mubar(x)}{2}\ ;\ e^{ -\frac{e\iE\beta^2}{4\tanh(e\iE s)}} \Biggl] \\ 
    &= -m\int_0^\infty \frac{\mathrm{d}s}{2\pi^{3/2}\beta s}\  e^{-sm^2} \sqrt{e\iE\coth(e\iE s)}\  e^{ -\frac{\mubar(x)^2}{e\iE\coth(e\iE s)}}   \\
    &\qquad\Theta_2\Biggl[ -i\mubar(x)\frac{2\pi\tanh(e\iE s)}{e\iE\beta}\ ;\ e^{ -\frac{4\pi^2\tanh(e\iE s)}{e\iE\beta^2}} \Biggl],
\end{split}
\end{equation}
where we expressed the $\Theta_4$ function in terms of a $\Theta_2$ function.
The second equality in Eq.~\eqref{eq:infvolchiral} is explicitly given by
\begin{equation}\label{eq:qcondmatsum}
\begin{split}
    &\langle \bar\psi(x) \psi(x) \rangle = -\frac{m}{2\pi^{3/2}\beta}\sum_{n\in\mathds{Z}}\int_0^\infty\frac{\mathrm{d}s}{s}\ e^{-m^2s}  
    \sqrt{e\iE\coth(e\iE s)}\exp\Biggl[ -\frac{(\omega_n-\mubar(x))^2}{e\iE\coth(e\iE s)} \Biggl],
\end{split}
\end{equation}
with $\omega_n=(2n+1)\frac{\pi}{\beta}$, $n\in\mathds{Z}$ denoting the fermionic Matsubara frequencies.

\begin{figure}[h]
\centering
\includegraphics[width=1.0\textwidth]{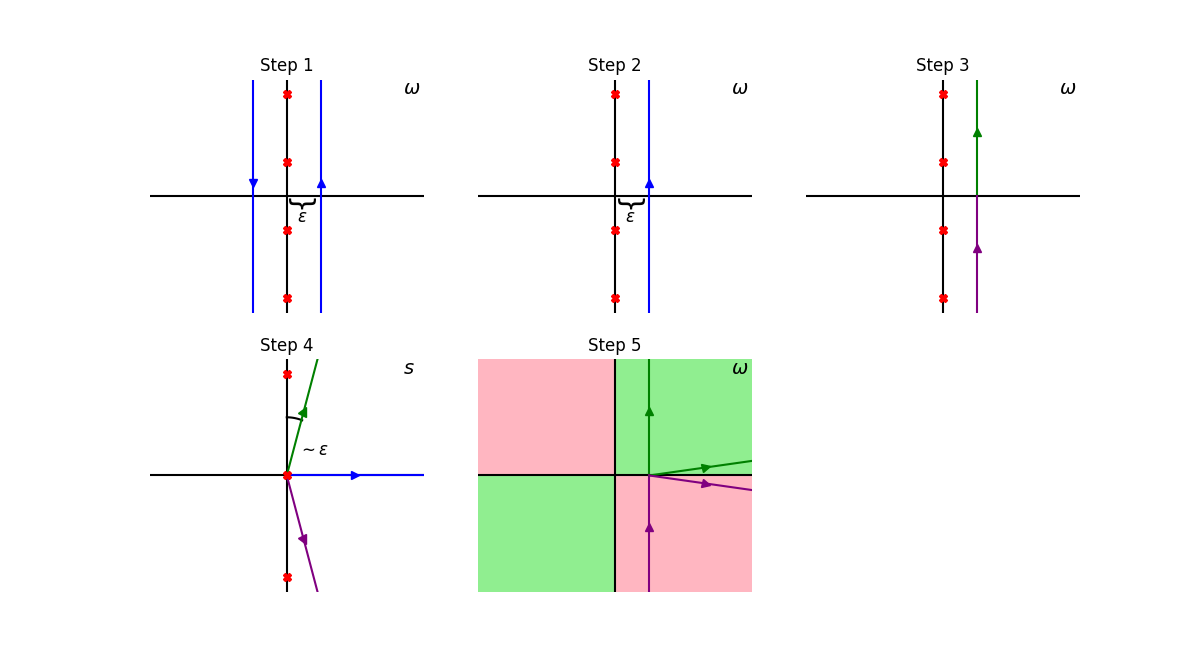}
\vspace{-1.5cm}
\caption{The contour integration required for the analytic continuation is shown in five steps. The red crosses symbolize the poles of the function, for which the contour integration is performed. 
}\label{fig:contourint}
\end{figure}

For the analytic continuation, we would like to find an expression in which the substitution $e\iE\rightarrow -ieE$ and $\mubar(x)\rightarrow -i\mubar(x)=-i(\mu-eEx_3)$ is possible.
First of all, Eq.~(\ref{eq:qcondmatsum}) can be expressed with an integration instead of a sum. This can be achieved by making use of the Fermi-Dirac distribution $\nf(\omega)=1/(e^{\beta\omega}+1)$, which has poles of residue $-1/\beta$ at $i\omega_n$ for all $n$, so a contour integration can be performed. This leads to the equation
\begin{equation}
    \langle \bar\psi(x) \psi(x) \rangle = -\frac{m}{2\pi^{3/2}}\int_0^\infty\frac{\mathrm{d}s}{s}\ e^{-m^2s} \sqrt{\coth(e\iE s)}
    \int_{\mathcal{C}} \frac{\mathrm{d}\omega }{2\pi i}\ \Bigl[ \frac{1}{2} - \nf(\omega+i\mubar(x)) \Bigl]\ \exp\Biggl[ \frac{\omega^2}{e\iE\coth(e\iE s)} \Biggl],
\end{equation}
where $\mathcal{C}$ is a contour from $-i\infty+\epsilon$ to $+i\infty+\epsilon$ plus from $+i\infty-\epsilon$ to $-i\infty-\epsilon$. This contour integration is shown in Fig.~\ref{fig:contourint} as the first step of the computation. The  factor $1/2$ can be inserted since it does not change the residue. But $1/2 - \nf(\omega)$ is an odd function, such that the path from $+i\infty-\epsilon$ to $-i\infty-\epsilon$ can be turned into a path from $-i\infty+\epsilon$ to $+i\infty+\epsilon$, shown in the second step of Fig.~\ref{fig:contourint}. This simplifies the local condensate to
\begin{equation}
\begin{split}
\langle \bar\psi(x) \psi(x) \rangle  
    &= -\frac{m}{2\pi^{3/2}}\int_0^\infty\frac{\mathrm{d}s}{s}\ e^{-m^2s} \sqrt{e\iE\coth(e\iE s)}
    \int_{-i\infty+\epsilon}^{+i\infty+\epsilon} 
    \frac{\mathrm{d}\omega }{2\pi i}\\
    &\quad\Biggl(
    \Bigl[ \frac{1}{2} - \nf(\omega+i\mubar(x)) \Bigl] - \Bigl[ \frac{1}{2} - \nf(-\omega+i\mubar(x)) \Bigl] \Biggl)\;
    \exp\Biggl[ \frac{\omega^2}{e\iE\coth(e\iE s)} \Biggl]\\
    &= -\frac{m}{2\pi^{3/2}}\int_0^\infty\frac{\mathrm{d}s}{s}\ e^{-m^2s} \sqrt{e\iE\coth(e\iE s)}
    \int_{-i\infty+\epsilon}^{+i\infty+\epsilon} 
    \frac{\mathrm{d}\omega }{2\pi i}\\
    &\quad\Bigl[ 1 - \nf(\omega+i\mubar(x))  -  \nf(\omega-i\mubar(x)) \Bigl] \;
    \exp\Biggl[ \frac{\omega^2}{e\iE\coth(e\iE s)} \Biggl].
\end{split}
\end{equation}

Next, it is possible to separate the vacuum contribution from the thermal contribution.
The vacuum contribution is
\begin{equation}
\begin{split}
    \langle \bar\psi(x) \psi(x) \rangle_{T=0} &= -\frac{m}{2\pi^{3/2}}\int_0^\infty\frac{\mathrm{d}s}{s}\ e^{-m^2s} \sqrt{e\iE\coth(e\iE s)}  
    \int_{-i\infty+\epsilon}^{+i\infty+\epsilon} 
    \frac{\mathrm{d}\omega }{2\pi i}  \exp\Biggl[ \frac{\omega^2}{e\iE\coth(e\iE s)} \Biggl]\\
    &= -\frac{m}{4\pi^2}\int_0^\infty\frac{\mathrm{d}s}{s}\ e^{-m^2s} e\iE\coth(e\iE s),
\end{split}
\end{equation}
where the Gaussian integral has been performed in the second step. Now, the analytic continuation $e\iE= -ieE$ can be performed, which leads to the vacuum contribution of the chiral condensate for real electric fields,\footnote{Notice that the resulting integrand has poles along the real $s$ axis. The integration path should be chosen such that it lies above the poles, since this gives a positive imaginary part for $\Leff$, being consistent with the vacuum decay for the Schwinger effect.}
\begin{equation}
    \langle \bar\psi(x) \psi(x) \rangle_{T=0} = -\frac{m}{4\pi^2}\int_0^\infty\frac{\mathrm{d}s}{s}\ e^{-m^2s} eE\cot(eEs).
\end{equation}

In turn, the thermal contribution is given by
\begin{equation}
    \langle \bar\psi(x) \psi(x) \rangle_{T\neq0}= \frac{m}{2\pi^{3/2}} \int_0^\infty\frac{\mathrm{d}s}{s}\ e^{-m^2s} \sqrt{e\iE\coth(e\iE s)} 
    \int_{-i\infty+\epsilon}^{+i\infty+\epsilon} 
    \frac{\mathrm{d}\omega }{2\pi i} \Nf(\omega)
    \exp\Biggl[ \frac{\omega^2}{e\iE\coth(e\iE s)} \Biggl],
\end{equation}
with the definition $\Nf(\omega)=\nf(\omega+i\mubar(x))  +  \nf(\omega-i\mubar(x))$. For the next steps, the proper time integration over $s$ and the integration over the frequencies $\omega$ need to be interchanged. Moreover, the frequency integration is split into two paths: one path from $i0+\epsilon$ to $i\infty+\epsilon$ and a second path from $-i\infty+\epsilon$ to $i0+\epsilon$, which are indicated respectively by the green and purple segments in the third step of Fig.~\ref{fig:contourint}. In the first (second) segment, the proper time integration is moved from the real axis to the positive (negative) imaginary axis, with a small real part $\epsilon$ for an angle.
This is shown in the fourth step of Fig.~\ref{fig:contourint} again with the segments indicated by green and purple. Altogether, we arrive at
\begin{equation}
\begin{split}
    \langle \bar\psi(x) \psi(x) \rangle_{T\neq0} = &\frac{m}{2\pi^{3/2}} \int_{i0+\epsilon}^{+i\infty+\epsilon} 
    \frac{\mathrm{d}\omega }{2\pi i}\ \Nf(\omega) 
    \int_0^\infty\frac{\mathrm{d}s}{s}\ e^{-m^2[i+\epsilon]s}  \\ &\sqrt{e\iE\coth(e\iE[i+\epsilon]s)}\exp\Biggl[ \frac{\omega^2}{e\iE\coth(e\iE[i+\epsilon]s)} \Biggl]\\ 
    &+ \frac{m}{2\pi^{3/2}} \int^{i0+\epsilon}_{-i\infty+\epsilon} 
    \frac{\mathrm{d}\omega }{2\pi i}\ \Nf(\omega) 
    \int_0^\infty\frac{\mathrm{d}s}{s}\ e^{-m^2[-i+\epsilon]s}  \\  &\sqrt{e\iE\coth(e\iE[-i+\epsilon]s)}\exp\Biggl[ \frac{\omega^2}{e\iE\coth(e\iE[-i+\epsilon]s)} \Biggl],
\end{split}
\end{equation}
which can now be analytically continued in $e\iE$. The replacement $e\iE\rightarrow -ieE$ leads to
\begin{equation}
\begin{split}
\langle \bar\psi(x) \psi(x) \rangle_{T\neq0} =
    & \frac{m}{2\pi^{3/2}} \int_{i0+\epsilon}^{+i\infty+\epsilon} 
    \frac{\mathrm{d}\omega }{2\pi i}\ \Nf(\omega) 
    \int_0^\infty\frac{\mathrm{d}s}{s}\ e^{-m^2[i+\epsilon]s} \\  &\sqrt{-ieE\coth(eE[1-i\epsilon]s)}\exp\Biggl[i \frac{\omega^2}{eE\coth(eE[1-i\epsilon]s)} \Biggl]\\ 
    &+ \frac{m}{2\pi^{3/2}} \int^{i0+\epsilon}_{-i\infty+\epsilon} 
    \frac{\mathrm{d}\omega }{2\pi i}\ \Nf(\omega) 
    \int_0^\infty\frac{\mathrm{d}s}{s}\ e^{-m^2[-i+\epsilon]s} \\  &\sqrt{ieE\coth(eE[1+i\epsilon]s)}\exp\Biggl[-i \frac{\omega^2}{eE\coth(eE[1+i\epsilon]s)} \Biggl]\,.
\end{split}
\end{equation}

The $\epsilon$ inside the $\coth$ function can be dropped, since it is not needed for the convergence of the proper time integration. Notice that $eE\coth(eEs)$ is positive such that it is possible to turn the first frequency path integration $\omega$ into an integration along the positive real axis with a small angle to its axis for convergence, which is equivalent to modifying  the denominator $\exp(i\omega^2/eE\coth(eEs))$ in the exponent by an $-i\epsilon$. Since the second path shows a $\exp(-i\omega^2/eE\coth(eEs))$, the same can be done here. The Gaussian functions only approach zero for $\omega\rightarrow\infty$ in two of the four quadrants of the complex plane. These quadrants are shown for the respective path in green and purple in the fifth step of Fig.~\ref{fig:contourint}, including the deformation of the corresponding paths.

This results in the expression
\begin{equation}
    \begin{split}
    \langle \bar\psi(x) &\psi(x) \rangle_{T\neq0} \\
    = &\frac{m}{2\pi^{3/2}} \int_{0}^{\infty} 
    \frac{\mathrm{d}\omega }{2\pi i}\ \Nf(\omega) 
    \int_0^\infty\frac{\mathrm{d}s}{s}\ e^{-m^2[+i+\epsilon]s}   \sqrt{-ieE\coth(eEs)}\exp\Biggl[+i \frac{\omega^2}{eE\coth(eEs)-i\epsilon} \Biggl]\\ 
    - &\frac{m}{2\pi^{3/2}} \int^{\infty}_{0} 
    \frac{\mathrm{d}\omega }{2\pi i}\ \Nf(\omega) 
    \int_0^\infty\frac{\mathrm{d}s}{s}\ e^{-m^2[-i+\epsilon]s}  \sqrt{+ieE\coth(eEs)}\exp\Biggl[-i \frac{\omega^2}{eE\coth(eEs)+i\epsilon} \Biggl]\,.
\end{split}
\end{equation}
Remember that $\Nf(\omega)$ implicitly depends on $\mubar(x)$.
The analytic continuation in $\mubar(x)$ is now possible, since the contour integration was moved away from the imaginary axis and it was a risk before, that real $\mubar(x)$ would cause a divergence. The substitution $\mubar(x)\rightarrow -i\mubar(x)=-i(\mu-eEx_3)$ leads to 
\begin{equation}
\begin{split}
    \langle \bar\psi(x) \psi(x) \rangle_{T\neq0} = &\frac{m}{\pi^{3/2}} \int_{0}^{\infty} 
    \frac{\mathrm{d}\omega }{2\pi } \Bigl[ \nf(\omega+\mubar(x))+\nf(\omega-\mubar(x)) \Bigl] \\ &\mathrm{Im}\Biggl(\int_0^\infty\frac{\mathrm{d}s}{s}\ e^{-m^2[i+\epsilon]s} \sqrt{eE\coth(eEs)}e^{-i\pi/4}\exp\Biggl[i \frac{\omega^2}{eE\coth(eEs)-i\epsilon} \Biggl]
    \Biggl),
\end{split}
\end{equation}
which, after the substitution $\omega\rightarrow-\omega$ in the second Fermi-Dirac distribution becomes 
\begin{equation}\label{thquarkend}
\begin{split}
    \langle \bar\psi(x) \psi(x) \rangle_{T\neq0} = &\frac{m}{\pi^{3/2}} \int_{-\infty}^{\infty} 
    \frac{\mathrm{d}\omega }{2\pi } \ff(\omega) \\ &\mathrm{Im}\Biggl(\int_0^\infty\frac{\mathrm{d}s}{s}\ e^{-m^2[i+\epsilon]s} \sqrt{eE\coth(eEs)}e^{-i\pi/4}\exp\Biggl[i \frac{\omega^2}{eE\coth(eEs)-i\epsilon} \Biggl]
    \Biggl),
\end{split}
\end{equation}
with the distribution $\ff(\omega)=\theta(\omega)\nf(\omega+\mubar(x))+\theta(-\omega)\nf(-\omega-\mubar(x))$.
This is precisely Eq.~(7) of Ref.~\cite{Elmfors:1994fw} and concludes the demonstration of consistency with~\cite{Elmfors:1994fw}.

\section{Weak field expansion of the oscillating field}
\label{app:4}

In this appendix, we give detailed account of the calculation to obtain our results for the Feynman diagrams in Eq.~(\ref{eq:Feynman}).

We start out with the calculation of 
\begin{equation}
     -\frac{\partial^2 \mathscr{L}_{\text{eff}}(x)}{\partial (eE)^2}\Biggl|_{E=\mu=0}
    = 
    \int\dd m
\begin{tikzpicture}[baseline=0.0cm]
    \filldraw [black] (0,0) circle (1.5pt);
    \begin{feynman}[small]
        \draw node at ($(-0.2cm,0.2cm)$) {$x$};
        \vertex at ($(0cm,0cm)$) (x);
        \vertex at ($(0.425cm,0.425cm)$) (y);
        \vertex at ($(0.425cm,-0.425cm)$) (z);
        \vertex at ($(0.425cm,0.85cm)$) (t1);
        \vertex at ($(0.425cm,-0.85cm)$) (t2);
    \diagram*{
        (x) --[fermion, quarter left] (y),
        (z) --[fermion, quarter left] (x),
        (y) --[fermion, half left] (z),
        (y) --[photon] (t1),
        (z) --[photon] (t2),
    };
    \end{feynman}
\end{tikzpicture}\Biggl|_{E=\mu=0}
\end{equation}
The Feynman rules imply
\begin{equation}
\begin{split}
    I = \begin{tikzpicture}[baseline=0.0cm]
    \filldraw [black] (0,0) circle (1.5pt);
    \begin{feynman}[small]
        \draw node at ($(-0.2cm,0.2cm)$) {$x$};
        \vertex at ($(0cm,0cm)$) (x);
        \vertex at ($(0.425cm,0.425cm)$) (y);
        \vertex at ($(0.425cm,-0.425cm)$) (z);
        \vertex at ($(0.425cm,0.85cm)$) (t1);
        \vertex at ($(0.425cm,-0.85cm)$) (t2);
    \diagram*{
        (x) --[fermion, quarter left] (y),
        (z) --[fermion, quarter left] (x),
        (y) --[fermion, half left] (z),
        (y) --[photon] (t1),
        (z) --[photon] (t2),
    };
    \end{feynman}
\end{tikzpicture}\Biggl|_{\mu=0}
=& \int\dd^4y\,\dd^4z\,\underset{R,P,Q}{\sumint} 2\frac{eA_0(y)}{eE}\,\frac{eA_0(x)}{eE}\, e^{iR\cdot(y-x)+iP\cdot(z-y)+iQ\cdot(x-z)}\\
&\mathrm{Tr}\Big[\frac{-1}{\slashed{R}-m}\gamma^0 \frac{-1}{\slashed{P}-m}\gamma^0 \frac{-1}{\slashed{Q}-m}\Big],
\end{split}
\end{equation}
where the contraction with the $\gamma$-matrices is in a  Minkowski sense and $P_0=i\omega$ holds for the fermionic Matsubara modes $\omega$. An additional factor of two appears due to the second derivative with respect to the electric field.

After simplifying the trace and performing the spatial integration we obtain the sum-integral
\begin{equation}
\begin{split}
    I = &\underset{P}{\sumint}\frac{4 m \left(2 k p \cos (\theta )+m^2+p^2-3 \omega ^2\right)}{k^2 \left(m^2+p^2+\omega ^2\right)^2 \left(k^2+2 k p \cos (\theta )+m^2+p^2+\omega ^2\right)} \\
    +&\underset{P}{\sumint}\cos(2kx_3) \frac{4 m \left(-k^2-m^2-P^2+4 \omega ^2\right)}{k^2 \left(m^2+P^2\right) \left(k^2-2 k p \cos (\theta )+m^2+P^2\right) \left(k^2+2 k p \cos (\theta )+m^2+P^2\right)}\\
    =& I_1 + \cos(2kx_3)\, I_2,
\end{split}
\end{equation}
where $\theta$ is the angle between the momenta $p$ and $k$, and $ P^2=\omega^2+p^2$ holds. Here we can directly observe the mixed representation nature of the diagram, revealed by the dependence of $I$ on the position $x_3$ as well as the momentum $k$.
In simplifying $I_1$ and $I_2$ we proceed in the same way, therefore we only discuss $I_1$ here.
By the usual contour integration involved in finite temperature calculations, we obtain the $T$-dependent part of the integration. The vacuum contribution to the susceptibility is not of interest here and would disappear anyway after the charge renormalization~\cite{Endrodi:2024cqn}. After an additional momentum shift, which is needed to get rid of the angular integration in the Fermi-Dirac distribution, we find
\begin{equation}
    I_1 = \int_0^\infty\dd p\, \frac{-p^2n_F(E_p)}{2\pi^2}\int_{-1}^{+1}\dd u\, \frac{m \left(3 k^3+k^2 p u+10 k m^2-2 k p^2 \left(u^2-5\right)+4 p u \left(m^2+p^2\right)\right)}{k^4E_p (k-2 p u) (k+p u) (k+2 p u) },
\end{equation}
where the substitution $\cos(\theta)=u$ was performed and we evaluate the expression in three dimensions.
We perform the angular integration $u$ by a principle value integration and solve for the mass integration (taking into account an integration constant such that the result vanishes for $m\to\infty$), resulting in
\begin{equation}
 -\frac{\partial^2 \mathscr{L}_{\text{eff}}(x)}{\partial (eE)^2}\Biggl|_{E=\mu=0}
    =\int\dd m\,I_1 = \int_0^\infty\dd p\,
    \frac{-p\, n_F(E_p) \left(\left(4 \left(m^2+p^2\right)-k^2\right) \log \left(\frac{k+2 p}{| k-2 p| }\right)+4 k p\right)}{4 \pi ^2 k^3E_p}.
\end{equation}

All other diagrams are treated by the same calculation steps with the additional modification of a possible chemical potential differentiation, which could also be represented by a $\gamma^0$ insertion on the level of the diagram. 

\section{Canonical susceptibility for a scalar field}
\label{app:5}

In this appendix we calculate the photon polarization diagram~\eqref{eq:Pimunudef} in scalar QED, which is needed to determine the electric and magnetic susceptibilities for charged pions, see Eq.~\eqref{eq:scalarchixi}. We use the Minkowski metric $g_{\mu\nu}=\textmd{diag}(1,-1,-1,-1)$.

For the susceptibilities, the second derivatives of the photon polarization tensor with respect to the momentum flowing through the diagram is required. This implies that we may ignore the contribution of the tadpole term to the polarization tensor (which is present in the scalar theory), since it is independent of the momentum. 
The relevant diagram is therefore given by 
\begin{equation}
    \Pi^{\mu\nu}(k) = 
    \begin{tikzpicture}[baseline=0.0]
\begin{feynman}[small]
\vertex at ($(0,0)$) (x);
\vertex at ($(0.85cm,0)$) (y);
\vertex at ($(x)-(0.85cm,0)$) (p);
\vertex at ($(y)+(0.85cm,0)$) (t);
\diagram*{
(x) --[fermion, half left, momentum = \(p+k\)] (y);
(y) --[fermion, half left, momentum = \(p\)] (x);
(p) --[photon, momentum = \(k\), edge label'=\(\mu\)] (x);
(y) --[photon,momentum = \(k\), edge label'=\(\nu\)] (t);
};
\end{feynman}
\end{tikzpicture}
    = (-ie)^2\int \frac{\mathrm{d}^4 p}{(2\pi)^4}(k+2p)^\mu (k+2p)^\nu \Sb(p) \Sb(k+p).
\end{equation}
The propagator in the real-time formulation for a spin-zero boson reads
\begin{equation}
    i \Sb(k)=\frac{i}{k^2-m^2+i\epsilon} + 2\pi \delta(k^2-m^2)\nb(|k^0|),
\end{equation}
with $\nb(\omega)=1/(e^{\beta\omega}-1)$ the Bose-Einstein distribution. 

For the electric susceptibility, the thermal part of the $\mu=\nu=0$ component is of interest. This is given by
\begin{equation}
    i\Pi_{00}^{T\neq 0}(k^0,\vec{k}) = 
    -2ie^2\int \frac{\mathrm{d}^4 p}{(2\pi)^4} (k^0+2p^0)^2\frac{\pi \nb(|p^0|)}{E_p} \frac{\delta(p^0-E_p)+\delta(p^0+E_p)}{(k+p)^2-m^2+i\epsilon}.
\end{equation}

Now the $p^0$ integration can be performed resulting in
\begin{equation}
\begin{split}
    i\Pi_{00}^{T\neq 0}(k^0,\vec{k}) =
    &-ie^2 \int \frac{\mathrm{d}^3 p}{(2\pi)^3}\frac{\nb(|E_p|)}{|E_p|} (k^0+2E_p)^2 \frac{1}{-2\vec{k}\cdot\vec{p}-\vec{k}^2+(k^0)^2+2k^0E_p+i\epsilon} \\ 
    &+\{E_p\rightarrow-E_p\},
\end{split}
\end{equation}
where $\{E_p\rightarrow-E_p\}$ represents a contribution of a flipped energy sign.

For the susceptibility -- being a static quantity -- $k^0$ needs to be taken to zero before taking $\vec{k}$ to zero. For $k^0=0$, the polarization tensor simplifies to
\begin{equation}
\begin{split}
    i\Pi_{00}^{T\neq 0}(0,\vec{k}) &=
    -2ie^2 \int \frac{\mathrm{d}^3 p}{(2\pi)^3}\frac{\nb(E_p)}{E_p} 4E_p^2 \frac{1}{-2\vec{k}\cdot\vec{p}-\vec{k}^2+i\epsilon}
     \\
    &= -8ie^2\int_0^\infty \frac{\mathrm{d} p}{(2\pi)^2}p^2\nb(E_p)E_p\int_{-1}^{+1}\mathrm{d}u \frac{1}{-2kpu-k^2+i\epsilon}
\end{split}
\end{equation}
where in the second step the angular integration has been performed and $k=|\vec{k}|$ holds. 

The integration over $u$ gives in the $\epsilon\rightarrow0$ limit the result $-\frac{1}{2kp}\ln\big(\frac{2p+k}{|2p-k|}\big)+i\frac{\pi}{2kp}\theta(k-2p)$. The imaginary part describes Landau damping, but this is not of interest here. See Ref.~\cite{Weldon:1982aq}, for a discussion of the imaginary part in the context of fermions.  
We continue to consider the real part of the photon polarization diagram, which reads
\begin{equation}
    i\text{Re}\Pi_{00}^{T\neq 0}(0,k)=
    i\frac{e^2}{\pi^2}\int_0^\infty\mathrm{d}p\, p\,\nb(E_p) E_p\frac{1}{k}\, \ln\Bigg( \frac{k+2p}{|k-2p|} \Bigg).
\end{equation}

Now taking the second derivative, one can make use of $\frac{\mathrm{d}k}{\mathrm{d}(k^2)} = \frac{1}{2k}$ since the photon polarization is an even function of $k$. This results in the principal value integral 
\begin{equation}\label{eq:xitrick}
\begin{split}
\xi&=
    \frac{1}{2e^2}\frac{\mathrm{d}^2 \text{Re}\Pi_{00}^{T\neq 0}(0,k) }{\mathrm{d}k^2} =  \frac{1}{e^2}\frac{\mathrm{d} \text{Re}\Pi_{00}^{T\neq 0}(0,k) }{\mathrm{d}(k^2)} \\ &= 
    \frac{1}{2k\pi^2} \,\mathrm{P}\int_0^\infty \mathrm{d}p\, p\,\nb(E_p) E_p \ \Bigg[ -\frac{1}{k^2}\ln\left( \frac{k+2p}{|k-2p|} \right) +\frac{1}{k}\Bigg( \frac{1}{k+2p} -\frac{1}{k-2p} \Bigg) \Bigg]\,,
\end{split}
\end{equation}
where $\mathrm{P}\int$ denotes principal value integration.

We note that naively interchanging the principal value integral and the $k\rightarrow0$ limit would result in 
\begin{equation}
\xi = 
    \frac{1}{12\pi^2} \,\mathrm{P}\int_0^\infty \mathrm{d}p\, \frac{\nb(E_p)E_p}{p^2},
\end{equation}
which is an IR-divergent integral. This step is clearly invalid. However, before taking the limit one can add the vanishing principal value integral~\cite{Endrodi:2022wym}
\begin{equation}
    0 =
    -\frac{4}{3\pi^2}\,\mathrm{P}\int_0^\infty \mathrm{d}p\, \nb(E_k)\, E_k \frac{p^2}{(k^2-4p^2)}\,,
\end{equation}
resulting in the IR-safe integral
\begin{equation}
    \xi= \frac{1}{12\pi^2}\int_0^\infty \mathrm{d}p\, \frac{E_p\nb(E_p) - m\nb(m)}{p^2}= \frac{1}{12\pi^2}\int_m^\infty\frac{\mathrm{d}\omega}{\sqrt{\omega^2-m^2}}\Big( \nb(\omega) +\omega\,n_B'(\omega) \Big),
\end{equation}
where in the second equality the substitution $\omega=E_p$ was used. This proves Eq.~\eqref{eq:resBxi} of the main text.

For the analogous calculation of the magnetic susceptibility, we need the spatial components of the photon polarization. For these, we have 
\begin{equation}
    i\frac{1}{2}\Pi_{jj}^{T\neq0}(k^0,\vec{k}) = 
    -ie^2\int\frac{\mathrm{d}^4p}{(2\pi)^4}(\vec{k}+2\vec{p})^2 \frac{\pi \nb(|p^0|)}{E_p} \frac{\delta(p^0-E_p)+\delta(p^0+E_p)}{(k+p)^2-m^2+i\epsilon},
\end{equation}
where once again the $p^0$ integration can be performed and the $k^0\rightarrow0$ limit can be taken. The notation $\Pi_{jj}^{T\neq0}$ implies a sum over the spatial components $j$. It follows, 
\begin{equation}
    i\frac{1}{2}\Pi_{jj}^{T\neq0}(0,k) = -ie^2\int_0^\infty \frac{\mathrm{d} p}{(2\pi)^2}p^2\frac{\nb(E_p)}{E_p}\int_{-1}^{+1}\mathrm{d}u \frac{k^2+4kpu+4p^2}{-2kpu-k^2+i\epsilon} ,
\end{equation}
for $k^0=0$. The $\epsilon\rightarrow0$ limit would also create an imaginary part, which we ignore again, resulting in
\begin{equation}
    i\frac{1}{2}\text{Re}\Pi_{jj}^{T\neq0}(0,k) = 
    -i\frac{e^2}{4\pi^2}\int_0^\infty \mathrm{d}p\, p^2\frac{\nb(E_p)}{E_p}\, \Bigg[ -4 +\frac{k^2-4p^2}{2kp}\ln\Bigg( \frac{k+2p}{|k-2p|} \Bigg) \Bigg].
\end{equation}
To find the second derivative with respect to $k$, one can make use of the same trick as in Eq.~\eqref{eq:xitrick} and we end up with
\begin{equation}
\begin{split}
    \frac{1}{4e^2}\frac{\mathrm{d}^2\text{Re}\Pi_{jj}^{T\neq0}(0,k)}{\mathrm{d}k^2}
    &=
    \frac{1}{2e^2}\frac{\mathrm{d}\text{Re}\Pi_{jj}^{T\neq0}(0,k)}{\mathrm{d}(k^2)} \\
    &= -\frac{1}{8\pi^2}\int_0^\infty\mathrm{d}p \,p^2\frac{\nb(E_p)}{E_p}\, \Bigg[ \frac{k^2+4p^2}{2k^3p}\ln\Bigg( \frac{k+2p}{|k-2p|} \Bigg) -\frac{1}{2k^2p} \Bigg]\,.
\end{split}
\end{equation}
For this expression, the $k\rightarrow0$ limit can be taken without any IR-divergence and we obtain
\begin{equation}
    \chi = 
    -\frac{1}{12\pi^2}\int_0^\infty \mathrm{d}p\, \frac{\nb(E_p)}{E_p}
    =
    -\frac{1}{12\pi^2}\int_m^\infty \frac{\mathrm{d}\omega}{\sqrt{\omega^2-m^2}}\,\nb(\omega),
\end{equation}
where the substitution $\omega=E_p$ was performed. This finally proves Eq.~\eqref{eq:resBchi} of the main text.

\bibliographystyle{JHEP.bst}
\bibliography{bib.bib}

\end{document}